\documentclass[11pt]{article}
\usepackage{float}
\usepackage{placeins}
\usepackage{amsmath}
\usepackage{amssymb}
\usepackage{amsthm}
\usepackage{amsfonts}
\usepackage{comment}
\usepackage{color}
\usepackage{mathrsfs}
\usepackage{braket}
\usepackage{graphicx}
\usepackage[subrefformat=parens]{subcaption}
\usepackage{cite}
\usepackage{here}
\usepackage{tikz}
\usetikzlibrary{decorations.pathmorphing}
\usepackage{caption}
\usepackage{multirow}
\usepackage[labelfont=bf]{caption}
\usepackage{array} 
\usepackage{hyperref}
\usepackage{listings}
\usepackage{graphicx}
\usepackage{multirow}
\usepackage{array}
\usepackage{geometry}
\usepackage{rotating}
\DeclareFontFamily{U}{stix2scr}{\skewchar\font127 }
\DeclareFontShape{U}{stix2scr}{m}{n} {<-> stix2-mathscr}{}
\DeclareMathAlphabet{\curly}{U}{stix2scr}{m}{n}
\SetMathAlphabet{\curly}{bold}{U}{stix2scr}{b}{n}
\DeclareFontShape{U}{stix2scr}{b}{n} {<-> stix2-mathscr-bold}{}
\DeclareFontShape{U}{stix2scr}{bx}{n} {<->ssub * stix2scr/b/n}{}

\renewcommand{\ell}{\curly{l}}

\usepackage[stable]{footmisc}

\makeatletter

\@addtoreset{equation}{section}
\makeatother

\begin{document}

\pagestyle{empty}

\title{\bf{Correspondence between quasinormal modes and grey-body factors of Schwarzschild--Tangherlini black holes}}

\date{}
\maketitle
\begin{center}
{\large 
Hyewon Han\footnote{dwhw101@dgu.ac.kr}, Bogeun Gwak\footnote{rasenis@dgu.ac.kr}
} \\
\vspace*{0.5cm}

{\it 
Department of Physics, Dongguk University, Seoul 04620, Republic of Korea
}

\end{center}

\vspace*{1.0cm}
\begin{abstract}
We investigate the correspondence between the quasinormal modes and grey-body factors of Schwarzschild--Tangherlini black holes. The gravitational perturbations in higher-dimensional black holes can be classified into scalar, vector, and tensor types. Considering the dimension-dependent forms of their effective potentials, the correspondence was examined for each dimension and perturbation mode. The accurate quasinormal modes were computed by suitably adopting the continued fraction and integration-through-midpoints methods, depending on the structure of the singularity. The grey-body factor can be obtained through its correspondence with the quasinormal mode, and its accuracy was analyzed by calculating its difference from the numerically computed grey-body factor. The correspondence failed for $l=2$ scalar gravitational perturbations in $D\ge7$ because the form of the potential is markedly different from that in four dimensions. The vector and tensor perturbation types exhibited good correspondence accuracies in all cases. The breakdown of the correspondence was rigorously demonstrated to stem from multiple potential barriers, and its applicability to each mode in higher dimensions was assessed.
\end{abstract}

\newpage
\baselineskip=18pt
\setcounter{page}{2}
\pagestyle{plain}
\baselineskip=18pt
\pagestyle{plain}
\setcounter{footnote}{0}

\hrule
\vspace{1em}

\tableofcontents

\vspace{3em}
\hrule

\vspace{1em}
\section{Introduction}
The black hole, which emerged as a solution to the Einstein field equations in general relativity, has a strong gravitational field. Any matter that enters a black hole cannot escape and is directed towards a curvature singularity at the center. Because a black hole hides internal events from external observers, its surface is known as an event horizon.
A spherically symmetric black hole is described by the Schwarzschild solution, which is characterized solely by mass \cite{Schwarzschild:1916uq}. A Kerr black hole, which is an axially symmetric black hole with mass and angular momentum, has a richer structure, such as a Cauchy horizon inside its event horizon and a ring singularity \cite{Kerr:1963ud}. Moreover, the event horizon is enclosed by a region in which a stationary observer rotates with the black hole, called an ergosphere. The rotational energy of a Kerr black hole can be extracted by considering a particle with negative energy that plunges into the event horizon within the ergosphere \cite{Penrose:1969pc}. This process is known as the Penrose process and leads to a decrease in the mass of the black hole. However, energy extraction can only be partially achieved because the irreducible mass, proportional to the square root of the surface area of the black hole, does not decrease during any classical process \cite{Christodoulou:1970wf,Christodoulou:1971pcn}. This is reminiscent of the irreducible nature of entropy. Bekenstein proposed that the entropy of a black hole is proportional to its surface area \cite{bekenstein2020black}, and the proportionality constant was later determined by Hawking \cite{hawking1975particle}. Based on these observations, one can expect that a black hole also has a temperature. Hawking explained that the quantum mechanical effects near the event horizon cause a black hole to emit thermal radiation at a temperature proportional to its surface gravity \cite{hawking1975particle}. These findings have laid the foundation for the study of black hole thermodynamics, which has provided insight into the quantum nature of black holes.

Quantum-gravity theories formulate quantum effects in the gravitational field. One of the candidate theories is string theory, which requires spacetime with more than four dimensions for mathematical consistency. In string theory, the derivation of the Bekenstein--Hawking entropy through several microstates was first obtained for a five-dimensional black hole \cite{Strominger:1996sh}. Higher-dimensional gravity is employed in brane-world models that address the problem of large scale difference between gravity and the weak force \cite{Arkani-Hamed:1998jmv,Randall:1999vf}. These models represent the observable universe as a four-dimensional brane embedded in a higher-dimensional bulk and explain that the Standard Model fields are confined to the brane, whereas gravity can propagate in the bulk with extra dimensions. 

In particular, string theory on the five-dimensional anti-de Sitter (AdS) space, which is a solution with a negative cosmological constant, is dual to the four-dimensional conformal field theory (CFT) \cite{Maldacena:1997re}. This conjecture is known as the AdS/CFT correspondence, which relates gravity theory in $(D+1)$ dimensions to $D$-dimensional gauge theory on the boundary \cite{Aharony:1999ti}. In this context, the correspondence between the thermodynamic properties of an AdS black hole and the CFT at the Hawking temperature was demonstrated \cite{Witten:1998qj}. It has prompted considerable interest in the analysis of higher-dimensional black holes and their dual interpretations. The underlying principle of the AdS/CFT correspondence led to the discovery of the de Sitter (dS)/CFT \cite{Strominger:2001pn} and Kerr/CFT correspondences \cite{Guica:2008mu}. 

A rotating black hole in AdS spacetime also serves as a key model in the description of scattering processes called superradiance \cite{Press:1972zz,Cardoso:2004nk}. If the frequency of a wave incident on a black hole satisfies certain conditions associated with the angular velocity of the black hole, the scattered wave is amplified. In this case, the AdS boundary acts as a mirror surrounding the black hole, repeatedly enhancing the reflected wave and resulting in the instability of the system \cite{Brito:2015oca,Mollicone:2024lxy,Carballo:2025ajx,Dorlis:2025amf}. The stability of a black hole was investigated by analyzing the complex frequency of the fields around it, which determined its amplification or decay. A decaying oscillation in response to the perturbations of test fields or gravitational perturbations is known as a quasinormal mode \cite{Konoplya:2011qq,Konoplya:2003ii,Zhidenko:2006rs,Gwak:2022nsi,BarraganAmado:2023wxt,Han:2024rus}. The real part of the associated frequency provides the oscillation frequency, and the imaginary part determines the damping time. The quasinormal mode describes the ringdown phase of gravitational waves emitted from astrophysical processes such as collisions of binary black holes \cite{Dreyer:2003bv,Berti:2005ys,Berti:2025hly}. Because the quasinormal frequency depends solely on the parameters of the final black hole, regardless of the initial state of the system, the intrinsic information about the black hole can be extracted from the observed ringdown data. Furthermore, the quasinormal frequency is used to determine the blueshift instability of the Cauchy horizon of the black hole resulting from competition with its surface gravity \cite{Cardoso:2017soq,Gwak:2018rba,Destounis:2020yav,Courty:2023rxk,Shao:2023qlt,Davey:2024xvd,Lin:2024beb} to validate the strong cosmic censorship conjecture \cite{Penrose:1969pc}.

A quasinormal mode is characterized by boundary conditions with ingoing modes on the event horizon and outgoing modes at spatial infinity. If outgoing modes at the horizon are allowed, a classical scattering problem is encountered. The outgoing wave from the black hole is scattered by the gravitational potential barrier between the event horizon and infinity; thus, only a fraction of it passes through the barrier and reaches an observer at infinity. The partial transmission of radiation describes the deviation of a black hole from a perfect blackbody \cite{hawking1975particle,Page:1976df}. Therefore, the transmission probability of a black hole is called the grey-body factor and is used to analyze the Hawking radiation and superradiance. The grey-body factor for gravitational perturbations of higher-dimensional static black holes was studied in \cite{Cornell:2005ux,Cardoso:2005vb,Creek:2006ia,Harmark:2007jy}. Moreover, recent studies have found that the spectral amplitude of the gravitational wave signal from a Kerr black hole can be modeled in terms of the grey-body factor \cite{Oshita:2022pkc,Oshita:2023cjz,Okabayashi:2024qbz}. Given that the grey-body factor is characterized by the parameters of black holes, it serves as a useful tool to extract the properties of astrophysical black holes as well as the quasinormal frequency. 

The quasinormal modes and grey-body factors of a black hole are distinct spectral quantities associated with different boundary conditions. Interestingly, the correspondence between them was formulated for spherically symmetric black holes in four dimensions \cite{Konoplya:2024lir} using the WKB approach \cite{Schutz:1985km,Iyer:1986np}. The grey-body factor can be analytically obtained using the fundamental quasinormal mode at the eikonal limit, where the multipole number is very large. The correspondence shows high accuracy, even beyond the eikonal limit, by including overtones. It has been actively applied and tested for rotating black holes \cite{Konoplya:2024vuj,Pedrotti:2025idg,Huang:2025rxx}, dS black holes \cite{Malik:2024cgb}, quantum-corrected black holes \cite{Skvortsova:2024msa}, regular black holes \cite{Heidari:2024bbd,Tang:2025mkk,Lutfuoglu:2025ohb,Shi:2025gst,Bolokhov:2025lnt,Ji:2025nlc,Malik:2025dxn,Dubinsky:2025nxv,Lutfuoglu:2025blw,Malik:2025qnr,Dubinsky:2025wns,Lutfuoglu:2025mqa}, wormholes \cite{Bolokhov:2024otn,Chakraborty:2025wgk}, dilaton black holes \cite{Dubinsky:2024vbn,Lutfuoglu:2025eik}, non-commutative black holes \cite{Fan:2025ead}, black holes immersed in galactic halos \cite{Hamil:2025pte,Lutfuoglu:2025kqp} and in dark fluids and string clouds \cite{Yan:2025pvp}, Bumblebee black holes \cite{AraujoFilho:2025hkm,Heidari:2025oop}, brane-projected black holes \cite{Dubinsky:2025ypj}, and various black holes in theories coupled with nonlinear electrodynamics \cite{Hamil:2025cms,Malik:2025erb,AraujoFilho:2025zzf}, conformal gravity \cite{Konoplya:2025mvj,Lutfuoglu:2025hjy}, and higher-curvature gravity \cite{Sajadi:2025kah,Lutfuoglu:2025ldc,Hamil:2025fbn}. The analytical expression of the grey-body factor in terms of the quasinormal mode was extended to a five-dimensional black hole in \cite{Han:2025cal}, and this expression exhibits good accuracy for all three types of gravitational perturbations. Its validity stems from the form of an effective potential that has a single positive barrier and asymptotically reaches constant values, thereby enabling the WKB approach to perform well. For the gravitational perturbation of black holes in theories incorporating higher curvature corrections, the correspondence does not hold because the eikonal regime is not properly described by the WKB approach \cite{Konoplya:2017lhs,Konoplya:2017wot,Konoplya:2020bxa,Konoplya:2025afm}.

In this work, we investigated the validity of the correspondence between the quasinormal modes and grey-body factors for a higher-dimensional black hole. We considered gravitational perturbations, which can be classified into scalar, vector, and tensor types in higher dimensions \cite{Kodama:2003jz}, thereby enabling us to analyze the correspondence, even for tensor-type perturbations that do not appear in the four dimensions. For higher-dimensional black holes, some cases exist in which the effective potential deviates from the standard form to which the WKB method is applied, depending on the perturbation type, number of dimensions, and multipole number. This implies that the correspondence can differ from what it is in four dimensions, and {\it{it can be invalid in particular modes or dimensions}}. Thus, the correspondence for gravitational perturbations in dimensions greater than five must be explored. We examined it for each case by employing a Schwarzschild--Tangherlini black hole \cite{tangherlini1963schwarzschild}. To obtain grey-body factors via the correspondence analytically, the quasinormal frequencies for the three types of perturbations can be computed by adopting the continued fraction method \cite{Leaver:1985ax} and its modified method  \cite{Rostworowski:2006bp}, which introduces some midpoints. We explicitly provide the numerical data for the fundamental mode and the first overtone when $D=6,7,8$. (We used the data for $D=5$ from a previous work \cite{Han:2025cal}.) Substituting the results into the analytical formula enabled us to determine the grey-body factors. We also found the accurate grey-body factors using a numerical method to compare them with the values obtained from the correspondence. Consequently, we show the accuracy of the correspondence for each perturbation mode and dimension and discuss the extent of its applicability in higher dimensions.

The remainder of this paper is organized as follows. Section $2$ presents the background related to higher-dimensional black holes and the equations for the three types of gravitational perturbations. Section $3$ reviews the analytical formula that connects the quasinormal modes and grey-body factors. In Section $4$, the continued fraction method is applied to compute the quasinormal modes. Sections $5$, $6$, and $7$ present our numerical results for quasinormal frequencies and use them to obtain the grey-body factors through the correspondence for scalar, vector, and tensor perturbations, respectively. We also discuss the validity of the correspondence by comparing the grey-body factors obtained using the analytical formula with those computed numerically. Finally, we summarize our results in Section $8$.

We use the metric signature $(-,+,+,+,\cdots)$ and the units $c=G_D=1$.

\section{Gravitational perturbations of Schwarzschild--Tangherlini black hole}
We examined the correspondence between the quasinormal modes and grey-body factors of black holes in higher dimensions, $D \ge 5$. Higher-dimensional black holes are solutions to the Einstein field equations obtained from the variation of the Einstein--Hilbert action, generalized to higher dimensions as
    \begin{align}
        I = \frac{1}{16\pi} \int d^D x \sqrt{-g} \, \mathcal{R},
    \end{align}
where $g$ is the determinant of the metric and $\mathcal{R}$ is the curvature scalar. We consider a spherically symmetric and asymptotically flat black hole, which can be described by the Schwarzschild--Tangherlini metric \cite{tangherlini1963schwarzschild}
    \begin{align} \label{metric}
        ds^2=-f(r) dt^2 +f^{-1}(r)dr^2+r^2 d \Omega_{\mu}^2 ,
    \end{align}
where $\mu=D-2$ and $d \Omega_{\mu}^2$ is the line element on the unit $\mu$-sphere. The metric function is expressed as
    \begin{align}
        f(r)=1-\frac{M}{r^{\mu-1}}.
    \end{align}
The total mass $M_B$ of the black hole is related to the mass parameter $M$ as
        \begin{align} 
            M_B = \frac{\mu\Omega_\mu}{16 \pi} M, 
        \end{align}
where $\Omega_\mu=\frac{2\pi^{(\mu+1)/2}}{\Gamma(\frac{\mu+1}{2})}$ is the area of the unit $\mu$-sphere. The position $r=r_H$ of the horizon of the black hole is determined by $f(r_H)=0$. The Schwarzschild--Tangherlini spacetime features a single horizon corresponding to the event horizon at $r_H=M^{\frac{1}{\mu-1}}$. Therefore, the boundaries of the stationary region outside the black hole are the event horizon $r=r_H$ and spatial infinity $r=\infty$.

We focused on gravitational perturbations in a fixed Schwarzschild--Tangherlini background. The Einstein field equations for the gravitational perturbations of maximally symmetric black holes in higher dimensions can be reduced to a single wavelike equation in a gauge-invariant formalism \cite{Kodama:2003jz}. In dimensions greater than four, these equations can be classified into three types. Scalar- and vector-type perturbations correspond to polar and axial perturbations in four dimensions, respectively, whereas tensor-type perturbations  constitute a new mode that appears only in higher dimensions. Furthermore, unlike the two types of perturbations in four dimensions that have identical quasinormal spectra, all three types of perturbations have different spectra in higher dimensions. Our work explores the relationship between the quasinormal modes and grey-body factors, covering all types of gravitational perturbations.

The perturbation equations for the scalar, vector, and tensor types can be written as the Schrödinger-like equation
\begin{align} \label{waveeq}
        \frac{d^2 \Psi}{dr_*^2}+\left( \omega^2 - V(r) \right) \Psi=0,
    \end{align}
where the radial tortoise coordinate $r_*(r)$ is defined as $dr_*=f(r)^{-1} dr$. The effective potential $V(r)=V_S(r)$ for the scalar-type perturbation is given by
    \begin{align} \label{VS}
        V_S(r)&= \frac{f(r) Q(r)}{16 r^2 H(r)^2} ,
    \end{align}
where 
    \begin{align}
        Q(r) &= \mu^4(\mu+1)^2 y^3 + \mu(\mu+1)\left[ 4(2\mu^2 - 3\mu + 4)m + \mu(\mu-2)(\mu-4)(\mu+1) \right] y^2 \nonumber \\
            & - 12\mu\left[ (\mu - 4)m + \mu(\mu+1)(\mu - 2) \right] m y + 16m^3 + 4\mu(\mu+2)m^2, \\
        H(r)&=m+\frac{\mu(\mu+1)}{2} y, \\
        m&=l(l+\mu-1)-\mu, \\
        y&=\left(\frac{r_H}{r}\right)^{\mu-1},
    \end{align}
and $l \ge 2$ is the multipole number. The effective potentials $V(r)=V_V(r)$ for the vector-type and $V(r)=V_T(r)$ for the tensor-type perturbations are given by
    \begin{align}
        V_V(r)&=\frac{f(r)}{r^2} \left[l(l+\mu-1) + \frac{\mu(\mu-2)}{4} -\frac{3\mu^2}{4}y \right], \label{VV} \\
        V_T(r)&=\frac{f(r)}{r^2} \left[l(l+\mu-1) + \frac{\mu(\mu-2)}{4} +\frac{\mu^2}{4}y \right], \label{VT}
    \end{align}
respectively. All three types of potentials vanish at the event horizon $r=r_H$ and infinity $r=\infty$. By solving the wavelike equation \eqref{waveeq} with potentials \eqref{VS}--\eqref{VT}, we can find quasinormal modes and grey-body factors for the gravitational perturbations of Schwarzschild--Tangherlini black holes and test the correspondence between them.

\section{Correspondence between quasinormal modes and grey-body factors}
The quasinormal modes and grey-body factors are the main characteristics that emerge when describing the radiation process around a black hole. We discuss the relationship between these spectral quantities in spacetime generalized to higher dimensions. A quasinormal mode describes a damped oscillation characterized by a complex quasinormal frequency $\omega=\omega_R +i\omega_I$. The real part $\omega_R$ corresponds to the oscillation frequency, and the imaginary part $\omega_I$ determines the decay rate of the field. The quasinormal modes of a black hole require a condition that allows only outgoing modes at infinity and only incoming modes at the event horizon. As the infinity $r=\infty$ and horizon $r=r_H$ correspond to $r_* \pm \infty$ in the tortoise coordinate system, respectively, the boundary conditions for the wave equation \eqref{waveeq} are expressed as
        \begin{align} \label{qnmbc}
            \Psi=\begin{cases}
                    \, e^{+ i \omega r_*}, & \quad\mbox{for } \,r_* = + \infty, \\
                    \, e^{- i \omega r_*} , & \quad \mbox{for } \, r_* = - \infty,
                \end{cases}
        \end{align}
where we assume that $\omega_R>0$ and the time dependence of the perturbation is $e^{-i \omega t}$.

The grey-body factor of the black hole is associated with different boundary conditions. We consider a scattering problem using the following wave equation.
    \begin{align} \label{waveeq2}
        \frac{d^2 \Psi}{dr_*^2}+\left( \Omega^2 - V(r) \right) \Psi=0.
    \end{align}
Here, $\Omega$ is the real frequency of the waves that undergo scattering. The grey-body factor measures the partial flux of radiation through the gravitational potential barrier of a black hole. Because this process has intrinsic symmetry, an identical transmission probability is obtained regardless of whether the wave originates from the black hole or enters from infinity. Thus, we obtain the following boundary conditions by permitting the ingoing modes at spatial infinity.
        \begin{align} 
            \Psi=\begin{cases}
                    \, e^{- i \Omega r_*} + R  \, e^{+i \Omega r_*}, & \quad\mbox{for } \,r_* = + \infty, \\
                    \, Te^{- i \Omega r_*} , & \quad \mbox{for } \, r_* = - \infty,
                \end{cases}
        \end{align} \label{scatteringbc}
where $R$ and $T$ denote the reflection and transmission coefficients, respectively. Then, the grey-body factor $\Gamma$ is defined as
    \begin{align} 
        \Gamma = |T|^2 = 1- |R|^2.
    \end{align}
Usually, it is obtained by numerically solving equation \eqref{waveeq2} or using a semianalytical technique, such as the WKB approach. This paper focuses on the method of computing grey-body factors through their relationship with the quasinormal modes.
    
The correspondence between the quasinormal modes and grey-body factors of four-dimensional spherically symmetric black holes was found using the WKB method \cite{Konoplya:2024lir}. This method matches the solutions from a series expansion of the effective potential near its peak to the asymptotic solutions that satisfy the boundary conditions at the event horizon and at infinity. It performs well in spacetime with an effective potential characterized by a single positive peak and constant asymptotics. When the WKB method properly describes the eikonal regime of a large multipole number $l \gg 1$, the grey-body factor can be obtained exactly in terms of the fundamental quasinormal frequency in the eikonal limit. At smaller $l$ it is approximate, and its accuracy can be improved by incorporating corrections that contain higher overtones.

The correspondence was established from the sixth-order WKB formula, given by
    \begin{align} \label{wkb}
        i \frac{\omega^2-V_0}{\sqrt{-2V_0''}}-\sum^{6}_{i=2}\Lambda_i=\mathcal{K},
    \end{align}
where $V_0$ is the maximum value of the potential, $V_0''$ is the second derivative of the potential at its maximum, and $\Lambda_i$ is the $i$-th order correction beyond the eikonal limit \cite{Iyer:1986np,Konoplya:2003ii}. For quasinormal modes $\omega_n$, the right-hand side of the formula takes the value $\mathcal{K}=n+ \frac{1}{2}$, where $n=0,1,2,\cdots$ is the overtone index. 

When \eqref{wkb} is applied to the scattering problem using Eqs. \eqref{waveeq2}--\eqref{scatteringbc}, the value of $\mathcal{K}$ is expressed as a function of the real frequency $\Omega$. In this case, the grey-body factor can be calculated as
    \begin{align} \label{gbf}
        \Gamma (\Omega)=\frac{1}{1+e^{2 \pi i \mathcal{K}}}.
    \end{align}
The sixth-order WKB formulas for the quasinormal modes and the scattering problem allow one to obtain an analytical expression.
    \begin{align}
         i\mathcal{K}=& \frac{\Omega^2 - \mathrm{Re}[\omega_0]^2}{4 \, \mathrm{Re}[\omega_0] \, \mathrm{Im}[\omega_0]} - \frac{\mathrm{Re}[\omega_0]-\mathrm{Re}[\omega_1]}{16 \, \mathrm{Im}[\omega_0]} \nonumber \\
        & +\frac{\Omega^2 - \mathrm{Re}[\omega_0]^2}{32 \, \mathrm{Re}[\omega_0] \, \mathrm{Im}[\omega_0]} \left(\frac{\left(\mathrm{Re}[\omega_0]-\mathrm{Re}[\omega_1]\right)^2}{4 \, \mathrm{Im}[\omega_0]^2} - \frac{3\mathrm{Im}[\omega_0]-\mathrm{Im}[\omega_1]}{3 \, \mathrm{Im}[\omega_0]} \right) \nonumber  \\
        &- \frac{\left(\Omega^2 - \mathrm{Re}[\omega_0]^2\right)^2}{16 \, \mathrm{Re}[\omega_0]^3 \, \mathrm{Im}[\omega_0]} \left(1+\frac{\mathrm{Re}[\omega_0]\left(\mathrm{Re}[\omega_0]-\mathrm{Re}[\omega_1]\right)}{4 \, \mathrm{Im}[\omega_0]^2} \right) \nonumber \\
        &+\frac{\left(\Omega^2 - \mathrm{Re}[\omega_0]^2\right)^3}{32 \, \mathrm{Re}[\omega_0]^5 \, \mathrm{Im}[\omega_0]} \left(1+\frac{\mathrm{Re}[\omega_0]\left(\mathrm{Re}[\omega_0]-\mathrm{Re}[\omega_1]\right)}{4 \, \mathrm{Im}[\omega_0]^2} \right. \nonumber \\
        & \left. \qquad \qquad \qquad + \mathrm{Re}[\omega_0]^2 \left( \frac{\left(\mathrm{Re}[\omega_0]-\mathrm{Re}[\omega_1]\right)^2}{16 \, \mathrm{Im}[\omega_0]^4} - \frac{3\mathrm{Im}[\omega_0]-\mathrm{Im}[\omega_1]}{12 \, \mathrm{Im}[\omega_0]} \right)\right) + \mathcal{O}(l^{-3}),\label{corr}
    \end{align}
where $\omega_0$ and $\omega_1$ denote the quasinormal frequencies of the fundamental mode $n=0$ and the first overtone $n=1$, respectively. Higher overtones are involved using the higher-order WKB formula, but they make very few corrections, whereas the fundamental mode and the first overtone provide the dominant contributions.

It has recently been shown that the correspondence provides a good approximation of grey-body factor for various four-dimensional models. Furthermore, it can be applied to five-dimensional spherically symmetric black holes \cite{Han:2025cal}. Here, we discuss the validity of the correspondence for gravitational perturbations in arbitrary higher dimensions. Considering low multipole numbers, we analytically obtain the grey-body factors by substituting quasinormal frequencies $\omega_0$ and $\omega_1$ into formula \eqref{corr} and compare them with the numerically calculated grey-body factors.

\section{Calculation of quasinormal mode}
To analyze the correspondence for the three types of gravitational perturbations of higher-dimensional black holes, we computed the fundamental quasinormal mode $\omega_0$ and the first overtone $\omega_1$ for each case. The continued fraction method was adopted to obtain accurate values. It was originally introduced by Leaver to calculate the quasinormal modes of Kerr black holes \cite{Leaver:1985ax} and has been extensively used for various purposes, including higher-dimensional black holes \cite{Ida:2002zk,Cardoso:2003vt,Cardoso:2003qd,Zhidenko:2006rs,Rostworowski:2006bp,Konoplya:2007jv,Lu:2023par}. This section reviews the continued fraction method to determine the fundamental mode and the first overtone for scalar-, vector-, and tensor-type gravitational perturbations.

We begin by constructing an ansatz for the solution to Eq. \eqref{waveeq}, satisfying the quasinormal boundary condition \eqref{qnmbc}. For even dimensions $D>4$, \cite{Rostworowski:2006bp}
    \begin{align} \label{ansatz1}
        \Psi(r)=e^{i \omega r} \left(\frac{r-r_H}{r}\right)^{-i \omega r_H/(D-3)} u(\rho(r)),  
    \end{align}
and for odd dimensions $D>4$,
    \begin{align} \label{ansatz2}
        \Psi(r)=e^{i \omega r} \left(\frac{r-r_H}{r+r_H}\right)^{-i \omega r_H/(D-3)} u(\rho(r)),  
    \end{align}
where
    \begin{align} \label{frob}
        u(\rho(r))= \sum^{\infty}_{k=0} a_k \rho^k=\sum^{\infty}_{k=0} a_k \left(\frac{r-r_H}{r}\right)^k,  
    \end{align}
is the Frobenius series. This series must converge in the region between the event horizon $\rho=0$ $(r=r_H)$ and infinity $\rho=1$ $(r=\infty)$. In other words, no additional singular points of Eq. \eqref{waveeq} may be present within the unit circle $|\rho|<1$. 

Upon substituting the ansatz \eqref{ansatz1}--\eqref{ansatz2} into Eq. \eqref{waveeq}, we obtain $N$-term recurrence relation for the coefficients $\{a_k\}$ of the series as
    \begin{align} \label{recur}
        \sum^{\min(N-1,k+1)}_{j=0}c^{(N)}_{j,k}(\omega)  \, a_{k+1-j}=0, \qquad \mathrm{for} \ k=0,1,2,\cdots,
    \end{align}
where we take $a_0=1$. The number $N$ of terms and coefficients $\{c^{(N)}_{j,k}\}$ are determined by each type of effective potential: \eqref{VS}, \eqref{VV}, and \eqref{VT}. For scalar-type perturbations, the recurrence relation contains $4(D-3)+1$ terms for even dimensions and $4(D-3)$ terms for odd dimensions. For vector- and tensor-type perturbations, one obtains the $(2(D-3)+1)$-term recurrence relation for even dimensions and the $(2(D-3))$-term recurrence relation for odd dimensions. 

The $N$-term recurrence relation \eqref{recur} can be reduced to a three-term recurrence relation via Gaussian elimination \cite{Zhidenko:2006rs,Konoplya:2011qq}. Its coefficients are obtained by performing the following process from $p=N-1$ to $p=3$.
    \begin{align} 
        c^{(p)}_{j,k}(\omega)&=c^{(p+1)}_{j,k}(\omega) \qquad \mathrm{for} \ j=0, \ \mathrm{or} \ (k+1) < p, \nonumber \\
        c^{(p)}_{j,k}(\omega)&=c^{(p+1)}_{j,k}(\omega) -\frac{c^{(p+1)}_{p,k}(\omega) \, c^{(p)}_{j-1,k-1}(\omega)}{c^{(p)}_{p-1,k-1}(\omega)}. \label{GaE}
    \end{align}
Subsequently, by writing $\alpha_k=c^{(3)}_{0, k}(\omega), \ \beta_k=c^{(3)}_{1, k} (\omega)$, and $\gamma_k=c^{(3)}_{2, k}(\omega)$, we have 
    \begin{align} 
        \alpha_0 & a_1+\beta_0 a_0 =0, \label{3re1} \\ 
        \alpha_k & a_{k+1}+\beta_k a_k + \gamma_k a_{k-1}=0, \qquad \mathrm{for} \ k>0, \label{3re2}
    \end{align}
The condition for the convergence of the series \eqref{frob} can be written in terms of an infinite continued fraction. Specifically,
    \begin{align} \label{n0}
        \frac{a_1}{a_0}&=-\frac{\beta_0}{\alpha_0}, \nonumber \\
        \frac{a_1}{a_0}&= \frac{- \gamma_1}{\beta_1-\frac{\alpha_1 \gamma_2}{\beta_2-\frac{\alpha_2 \gamma_3}{\beta_3- \cdots}}} \equiv \frac{-\gamma_1}{\beta_1 -}\frac{\alpha_1 \gamma_2}{\beta_2 -}\frac{\alpha_2 \gamma_3}{\beta_3 -} \dots.
    \end{align}
By equating the right-hand sides of \eqref{n0} and solving the equation, we obtain the quasinormal frequency $\omega_0$ of the fundamental mode $n=0$. The higher overtones $\omega_n$ can be obtained by inverting the equation as
    \begin{align} 
        \beta_n-\frac{\alpha_{n-1}\gamma_n}{\beta_{n-1} -}\frac{\alpha_{n-2}\gamma_{n-1}}{\beta_{n-2} -}\cdots\frac{\alpha_0 \gamma_1}{\beta_0}=\frac{\alpha_n \gamma_{n+1}}{\beta_{n+1} -}\frac{\alpha_{n+1} \gamma_{n+2}}{\beta_{n+2} -}\frac{\alpha_{n+2} \gamma_{n+3}}{\beta_{n+3} -} \cdots, \quad \mathrm{for}\ n \ge1 ,
    \end{align}
and determine its most stable roots. To enhance the convergence of the infinite continued fraction, we employed the Nollert improvement \cite{Nollert:1993zz, Zhidenko:2006rs}, a technique for truncating the continued fraction at a large depth and replacing its asymptotic tail with an analytical approximation.

The continued fraction method relies on the convergence of the Frobenius series \eqref{frob}. If any singularity exists inside a unit circle $0<|\rho|<1$, the radius of convergence of the series $u(\rho)$ becomes smaller. Thus, the series does not converge at spatial infinity $\rho=1$. In this case, the original continued fraction method breaks down and we apply the method of integration through the midpoints developed in \cite{Rostworowski:2006bp}. By introducing a regular midpoint $\rho=\rho_0$ within the radius of convergence such that $\rho=1$ becomes its nearest singularity, the series $u(\rho)$ can be continued analytically.

We construct a new series around the midpoint $\rho_0$ as follows: 
    \begin{align} \label{news}
        u(\rho)=\sum^{\infty}_{k=0} a_k \rho^k=\sum^{\infty}_{k=0} \tilde{a}_k (\rho-\rho_0)^k,
    \end{align}
where 
    \begin{align} \label{ta01}
        \tilde{a}_0=\sum^{\infty}_{k=0} a_k \rho_0^k, \qquad \tilde{a}_1=\sum^{\infty}_{k=1} k a_k \rho_0^{k-1},
    \end{align}
The new series yields an $(N+1)$-term recurrence relation for coefficients $\{\tilde{a}_k\}$. Similar to the case of the initial series, the number of terms is reduced to three through Gaussian elimination \eqref{GaE}. Consequently, we obtain
    \begin{align} 
        \tilde{\alpha}_k & \tilde{a}_{k+1}+\tilde{\beta}_k \tilde{a}_k + \tilde{\gamma}_k \tilde{a}_{k-1}=0, \qquad \mathrm{for} \ k>0.
    \end{align}
This recurrence relation leads to a new, continued fraction.
\begin{align} \label{tn0}
        \frac{\tilde{a}_1}{\tilde{a}_0}= \frac{-\tilde{\gamma}_1}{\tilde{\beta}_1 -}\frac{\tilde{\alpha}_1 \tilde{\gamma}_2}{\tilde{\beta}_2 -}\frac{\tilde{\alpha}_2 \tilde{\gamma}_3}{\tilde{\beta}_3 -} \cdots.
    \end{align}
We find $\{a_k\}$ using the original recurrence relation \eqref{3re1}--\eqref{3re2} and substitute them into Eqs. \eqref{ta01} to determine $\tilde{a}_1/\tilde{a}_0$. Then, we solve Eq. \eqref{tn0} to find the eigenfrequency $\omega_0$. The overtone $\omega_n$ can be obtained from the inversion of the equation. 

If $\rho=1$ is not the nearest singularity to $\rho=\rho_0$, we iteratively construct the series for the next midpoints $\rho_1,\rho_2,$ and $\rho_3, \cdots,$ until the point $\rho=1$ lies within the radius of convergence. 
    \begin{align} \label{mid2}
        u(\rho)=\sum^{\infty}_{k=0} a_k \rho^k=\sum^{\infty}_{k=0} \tilde{a}_k (\rho-\rho_0)^k=\sum^{\infty}_{k=0} \bar{a}_k (\rho-\rho_1)^k=\cdots,
    \end{align}

We computed the quasinormal modes for the gravitational perturbations of a Schwarzschild--Tangherlini black hole by using the continued fraction method. Because the structure of the singularities of the wave equation \eqref{waveeq} differs depending on the type of perturbation, dimension $D$, and multipole number $l$, we adopted the appropriate approach in each case to ensure the convergence of the Frobenius series. The following sections present the numerical results and use them to examine the relationship between the quasinormal mode and grey-body factor.

\section{Correspondence for scalar gravitational perturbations}

We first analyze the correspondence between the quasinormal modes and grey-body factors for scalar-type gravitational perturbations. The effective potential for the three types of gravitational perturbations of the Schwarzschild--Tangherlini spacetime has a centrifugal form $ \sim f(r) \, l(l+1)/r^2$ in the eikonal limit. Hence, the correspondence established by the WKB approach is exact in the eikonal limit. It is approximate beyond the eikonal limit because of the asymptotic convergence of the WKB series. Therefore, we evaluated the accuracy of the correspondence for some of the lowest multipole numbers, $l=2,3,4$.

We numerically determined the fundamental mode $\omega_0$ and the first overtone $\omega_1$ for $D=6,7,8$ by using the continued fraction method. By substituting these values into the correspondence formula \eqref{corr}, we obtained the grey-body factors for the real frequency $\Omega$ of a wave undergoing scattering. To verify the accuracy of the results, the differences between the grey-body factors obtained by the correspondence and those obtained by numerically solving the wave equation \eqref{waveeq2} were computed. We employed the Mathematica code \texttt{GrayHawk} \cite{Calza:2025whq}, originally designed to calculate the grey-body factors of four-dimensional spherically symmetric black holes, and extended it to higher dimensions.

\subsection{Correspondence in $4<D\le6$}

We present the numerical results $r_H \omega_n$ for the fundamental mode $n=0$ and the first overtone $n=1$. For scalar-type gravitational perturbations, the wave equation \eqref{waveeq} possesses singular points at $|\rho|<1$ depending on the dimension number $D$ and multipole number $l$. At $D=6$, the effective potential $V_S$ \eqref{VS} for $l=2$ has additional singular points at $|\rho| \simeq 0.93$, the roots of the function $H(r)$. This prevents the series \eqref{frob} from converging at $\rho=1$. We chose a regular point $\rho_0=0.5$ inside the radius of convergence, so that $\rho=1$ became its nearest singularity, and constructed a new series \eqref{news}. For $l=3,4$, all singular points satisfied $|\rho| >1$, and the original setting of the continued fraction method was applied. The results for $D=6$ are shown in \autoref{qnmS6}. Here and in the following sections, we will use the values presented in a previous work \cite{Han:2025cal} for the accurate quasinormal modes in $D=5$.

\begin{table}[h!]
\centering
\begin{tabular}{
|>{\centering\arraybackslash}p{0.5cm}|
 >{\centering\arraybackslash}p{3.5cm}|
 >{\centering\arraybackslash}p{3.5cm}|
 >{\centering\arraybackslash}p{3.5cm}| }
 
\hline\hline
\multicolumn{4}{|c|}{\textbf{$D=6$}} \\
\hline\hline

$n$ & $l=2$ & $l=3$ & $l=4$ \\
\hline
$0$   & $1.136904 - 0.303576  i$ & $1.923054 - 0.399946 i$ & $2.623066 - 0.437897 i
$ \\
$1$   & $1.027190 - 1.005591 i $ & $1.766426 - 1.233715 i$ & $2.481683 - 1.333891 i$ \\
\hline\hline

\end{tabular}
    \caption{Quasinormal frequency $r_H \omega_n$ for scalar gravitational perturbations in $D=6$} 
\label{qnmS6}
\end{table}

\begin{figure}[h!]
\noindent\begin{subfigure}[b]{0.5\textwidth}
    \centering
    \includegraphics[scale=0.42]{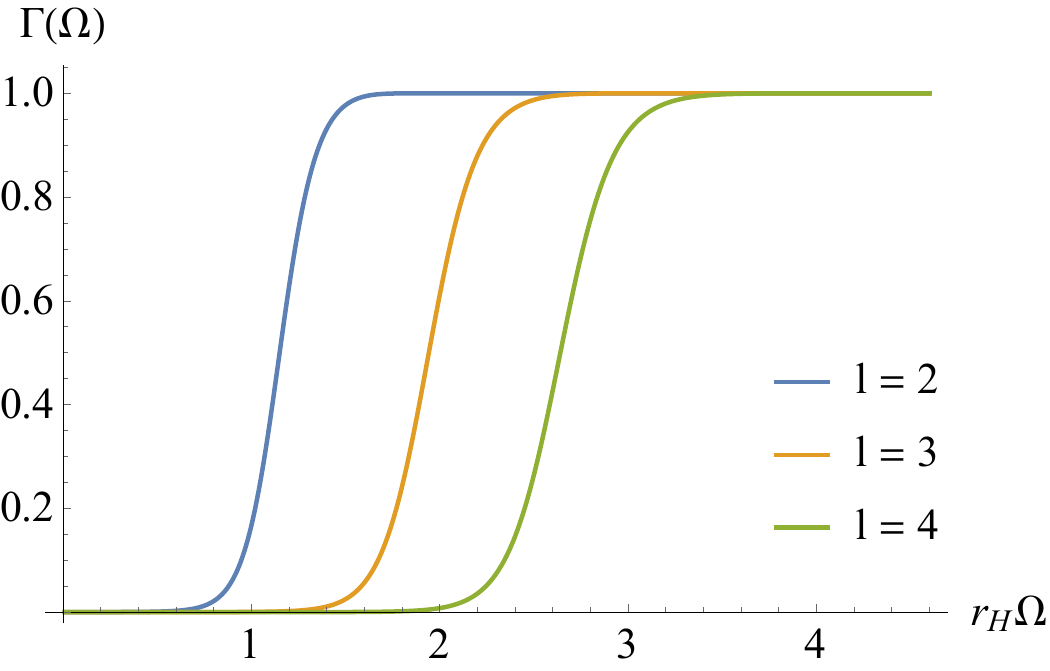}
\end{subfigure}%
\noindent\begin{subfigure}[b]{0.5\textwidth}
    \centering
    \includegraphics[scale=0.42]{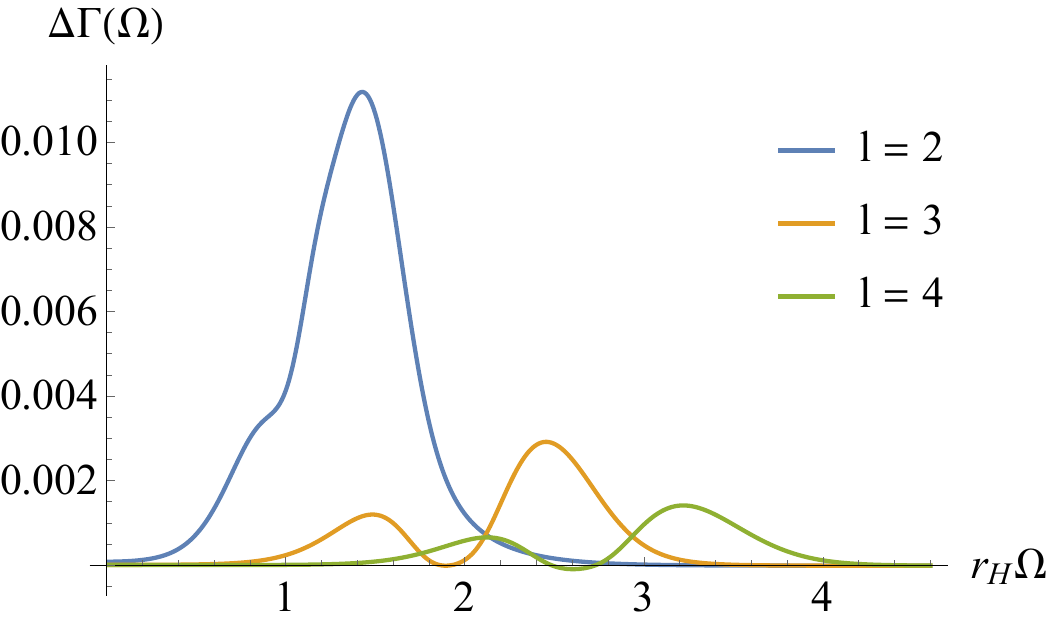}
\end{subfigure}
\caption{Left: Grey-body factors obtained by the correspondence with the quasinormal modes for scalar gravitational perturbations for $l=2$ (blue), $l=3$ (yellow), and $l=4$ (green) in $D=6$. Right: Differences between the grey-body factors calculated by using the correspondence and the numerical method.} \label{fig6s}
\end{figure}

It was demonstrated that the correspondence yields the grey-body factor with good accuracy for $l=2,3,4$ modes of scalar gravitational perturbations of the five-dimensional black hole \cite{Han:2025cal}. To test the correspondence for $D=6$, we obtained the grey-body factor \eqref{gbf} by substituting the quasinormal frequencies into the analytical formula \eqref{corr} and plotted them with respect to the real frequency $r_H \Omega$ in the left panel of \autoref{fig6s}. The right panel shows the difference between the approximate values determined by the correspondence and the accurate values obtained by numerical calculation. 

The small differences $\Delta \Gamma (\Omega)$ in \autoref{fig6s} indicate that the relationship between the quasinormal modes and grey-body factors for scalar-type perturbations holds at low multipole numbers in $D=6$. Furthermore, the difference decreases as $l$ increases, which is consistent with the fact that the correspondence is precise at high multipole numbers.

\subsection{Breakdown : $D\ge7$}

We examined the correspondence for dimensions greater than six. In $D \ge7$, additional singularities at $|\rho|<1$ appear in the effective potential $V_S$ \eqref{VS} for $l=2,3,4$. Consequently, the Frobenius series must be continued analytically through a positive real midpoint $\rho_0$. For $l=2$ in $D=8$, two midpoints are required to ensure that the series converges to the point $\rho=1$. In this case, we employed $\rho_0=0.4$ and $\rho_1=0.7$ to construct the series \eqref{mid2} and to obtain an equation with an infinite continued fraction. By solving these equations, we determined the quasinormal frequencies $r_H \omega_n$ of the scalar-type perturbation in $D=7,8$, as presented in \autoref{qnmS}. \footnote{Our numerical results for the $l=2$ fundamental mode $n=0$ in $D=6, 7$ reproduce the data obtained by using the continued fraction method in \cite{Konoplya:2007jv}. We performed the calculations for mode $n=1$ and $D=8$ by applying the same approach.}

\begin{table}[h!]
\centering
\begin{tabular}{
|>{\centering\arraybackslash}p{0.5cm}|
 >{\centering\arraybackslash}p{3.5cm}|
 >{\centering\arraybackslash}p{3.5cm}|
 >{\centering\arraybackslash}p{3.5cm}| }
 
\hline\hline
\multicolumn{4}{|c|}{\textbf{$D=7$}} \\
\hline\hline

$n$ & $l=2$ & $l=3$ & $l=4$ \\
\hline
$0$   & $ 1.339158-0.400860i $ & $ 2.232413-0.489720i $ & $3.011185-0.533841 i
$\\
 $1$   & $1.185695-1.100378 i $ & $1.973249-1.464668i$ & $2.785053-1.617320i$\\
\hline\hline

\multicolumn{4}{|c|}{\textbf{$D=8$}} \\
\hline\hline

$n$ & $l=2$ & $l=3$ & $l=4$ \\
\hline
$0$   & $1.563879 - 0.603123 i$ & $2.557452-0.599286 i$ & $3.391730-0.630915i
$\\
 $1$   & $1.333951-1.054126i$ & $2.125659-1.651256i$ & $3.043718-1.872784i$\\
\hline\hline

\end{tabular}
    \caption{Quasinormal frequency $r_H \omega_n$ for scalar gravitational perturbations in $D=7,8$} 
\label{qnmS}
\end{table}

\begin{figure}[h!]
\noindent\begin{subfigure}[b]{0.5\textwidth}
    \centering
    \includegraphics[scale=0.42]{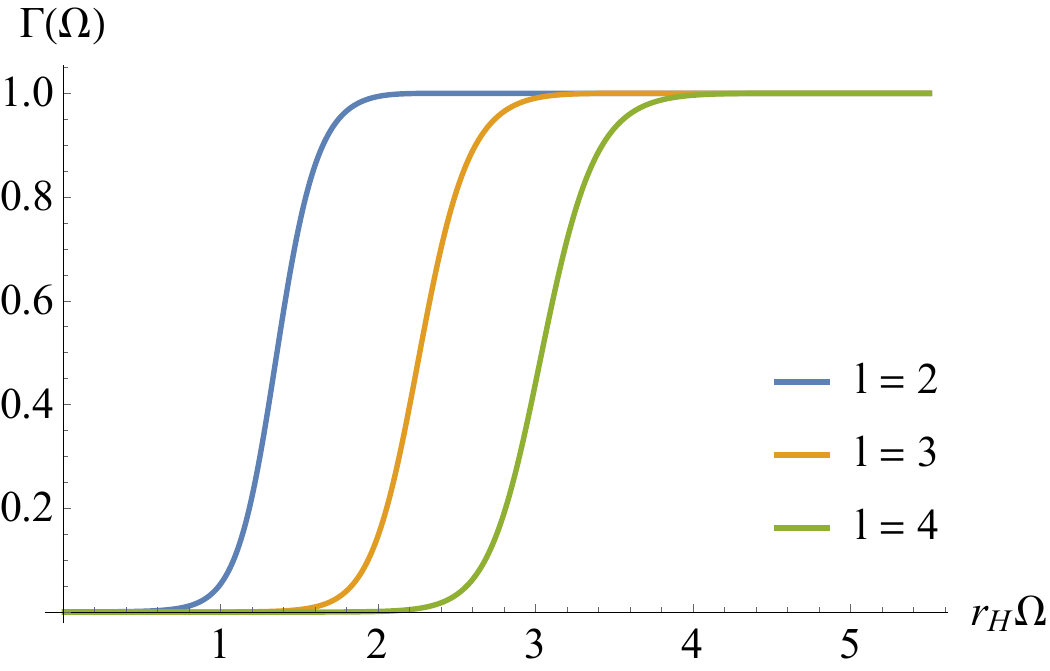}
\end{subfigure}%
\noindent\begin{subfigure}[b]{0.5\textwidth}
    \centering
    \includegraphics[scale=0.42]{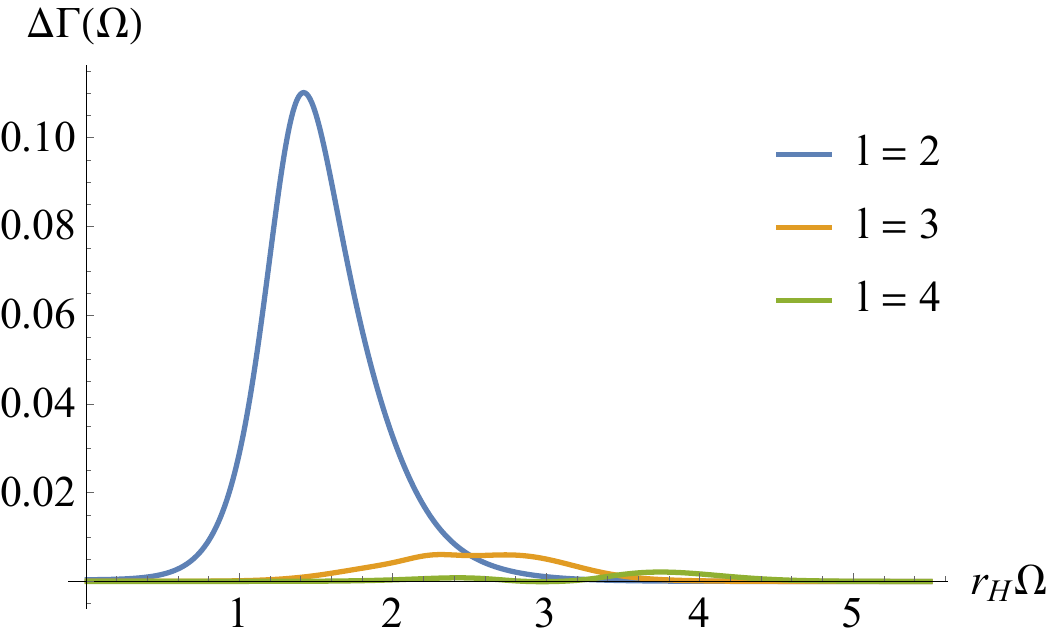}
\end{subfigure}
\caption{Left: Grey-body factors obtained by using the correspondence with the quasinormal modes for scalar gravitational perturbations for $l=2$ (blue), $l=3$ (yellow), and $l=4$ (green) in $D=7$. Right: Differences between the grey-body factors calculated by using the correspondence and the numerical method.} \label{fig7s}
\end{figure}

\begin{figure}[h!] 
\noindent\begin{subfigure}[b]{0.5\textwidth}
    \centering
    \includegraphics[scale=0.38]{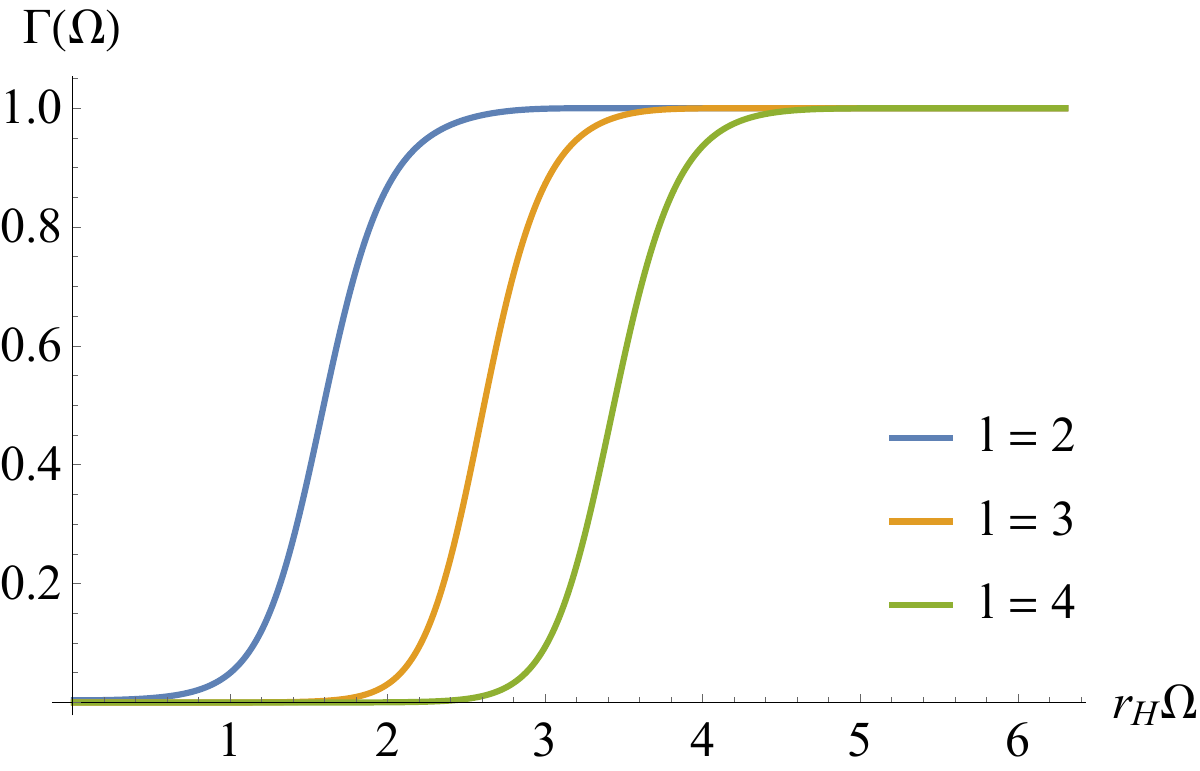}
\end{subfigure}%
\noindent\begin{subfigure}[b]{0.5\textwidth}
    \centering
    \includegraphics[scale=0.38]{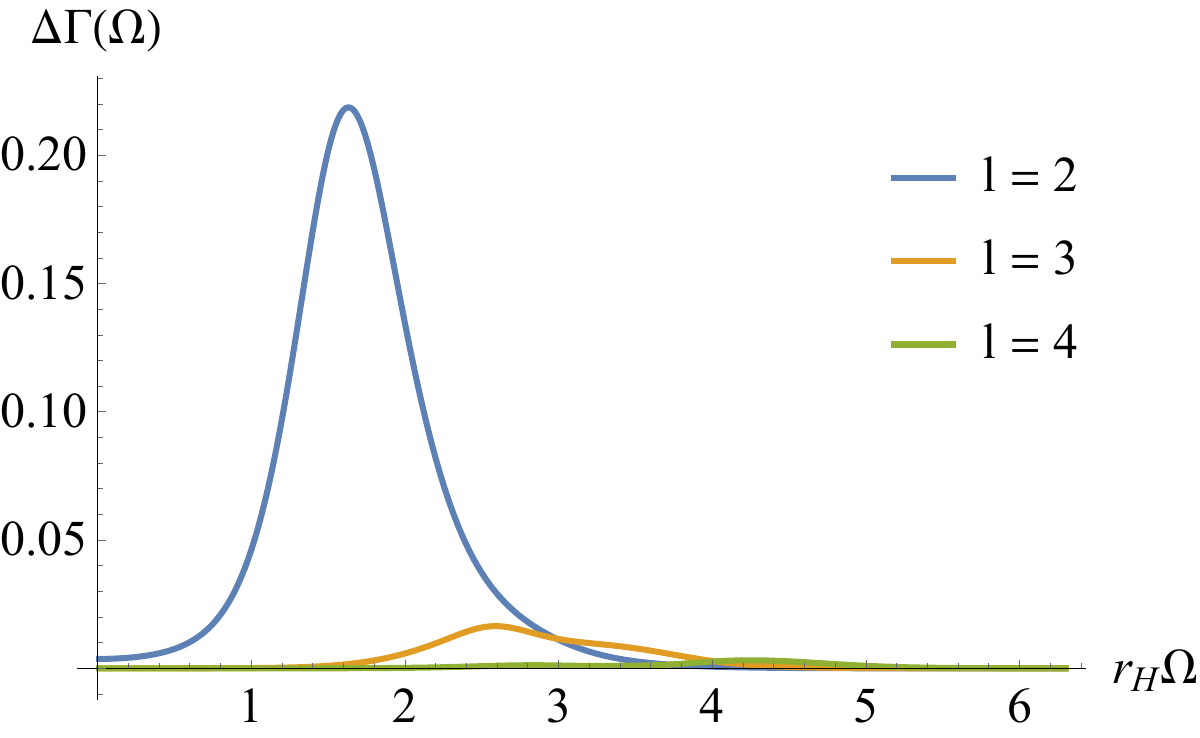}
\end{subfigure}
\caption{Left: Grey-body factors obtained by using the correspondence with the quasinormal modes for scalar gravitational perturbations for $l=2$ (blue), $l=3$ (yellow), and $l=4$ (green) in $D=8$. Right: Differences between the grey-body factors calculated by using the correspondence and the numerical method.} \label{fig8s}
\end{figure}

The results can be directly substituted into Eq. \eqref{corr} to determine the grey-body factors $\Gamma(\Omega)$. \autoref{fig7s} and \autoref{fig8s} show the grey-body factors obtained from the correspondence with the quasinormal modes in $D=7$ and $D=8$, respectively. The panels on the right illustrate how they differ from the numerically calculated grey-body factors. 

\begin{figure}[h!]
\noindent\begin{subfigure}[b]{0.33\textwidth}
    \centering
    \includegraphics[scale=0.55]{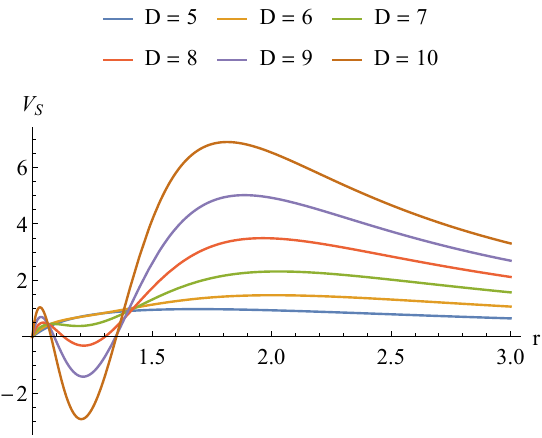}
\end{subfigure}%
\noindent\begin{subfigure}[b]{0.33\textwidth}
    \centering
    \includegraphics[scale=0.55]{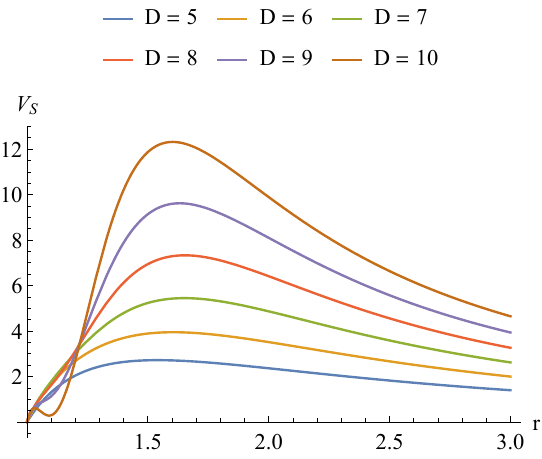}
\end{subfigure}%
\noindent\begin{subfigure}[b]{0.33\textwidth}
    \centering
    \includegraphics[scale=0.55]{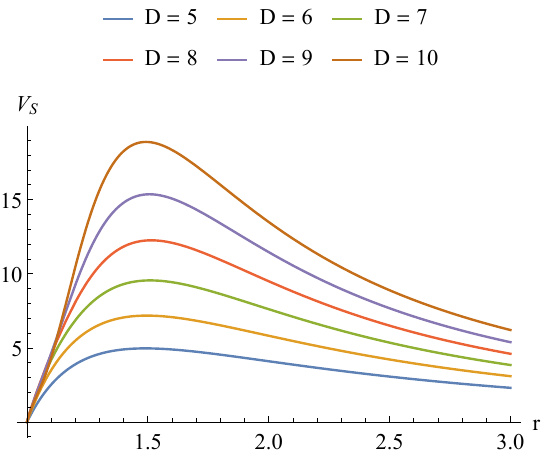}
\end{subfigure}%
\caption{ Effective potentials for scalar-type of gravitational perturbations of Schwarzschild--Tangherlini black holes ($r_H=1$) for $l=2$ (left), $l=3$ (center), and $l=4$ (right).} \label{VSp}\end{figure}

For both $D=7$ and $D=8$, the differences $\Delta\Gamma(\Omega)$ between the grey-body factors determined by using the correspondence and the accurate grey-body factors for $l=3, 4$ are small, suggesting the validity of the correspondence with reasonable accuracy. However, for the lowest multipole number, $l=2$, the differences plotted by the blue lines grow significantly, and thus, the correspondence fails. This behavior is due to the form of the effective potential $V_S$ \eqref{VS} for scalar-type perturbations of Schwarzschild--Tangherlini black holes. We plotted the potential outside the event horizon for $D=5,6,7,\cdots,10$ in \autoref{VSp}. The WKB approach used to formulate the correspondence was applied when the effective potential exhibited a single positive peak. For $l=2$, double peaks first appear in seven dimensions, represented by the green line in the left panel of \autoref{VSp}, and the second peak near the event horizon grows as the number of dimensions $D$ increases. In these cases with multiple barriers, the WKB approach is not applicable; therefore, the relationship between the quasinormal modes and grey-body factors breaks down. For the $l=3$ and $l=4$ modes, the effective potentials exhibit multiple peaks in $D \ge 10$ and $D \ge 12$, respectively. Consequently, we expect that the correspondence for the scalar gravitational perturbations will exhibit poor accuracy for arbitrary $D \ge 7$ for $l=2$, $D \ge 10$ for $l=3$, and $D \ge 12$ for $l=4$. 

\begin{figure}[h!]
\noindent\begin{subfigure}[b]{0.5\textwidth}
    \centering
    \includegraphics[scale=0.47]{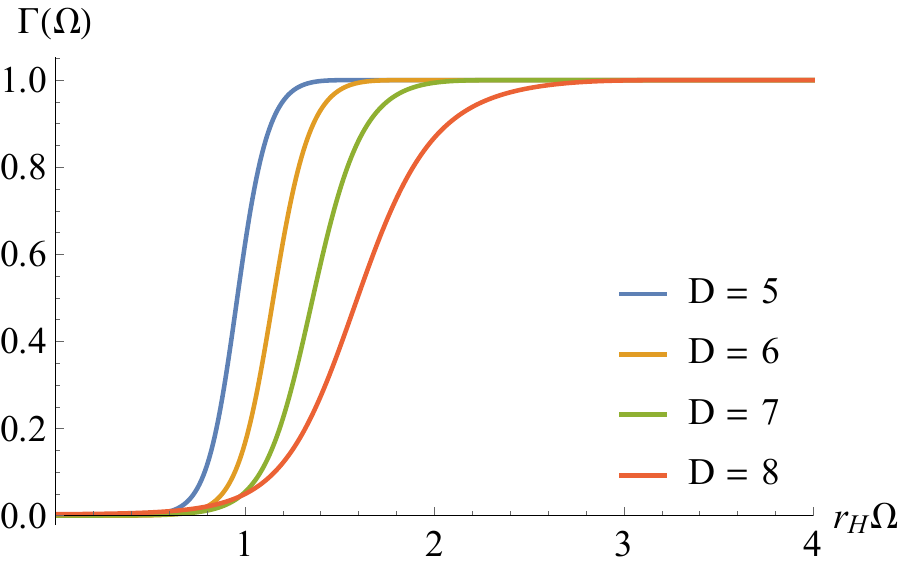}
\end{subfigure}%
\noindent\begin{subfigure}[b]{0.5\textwidth}
    \centering
    \includegraphics[scale=0.47]{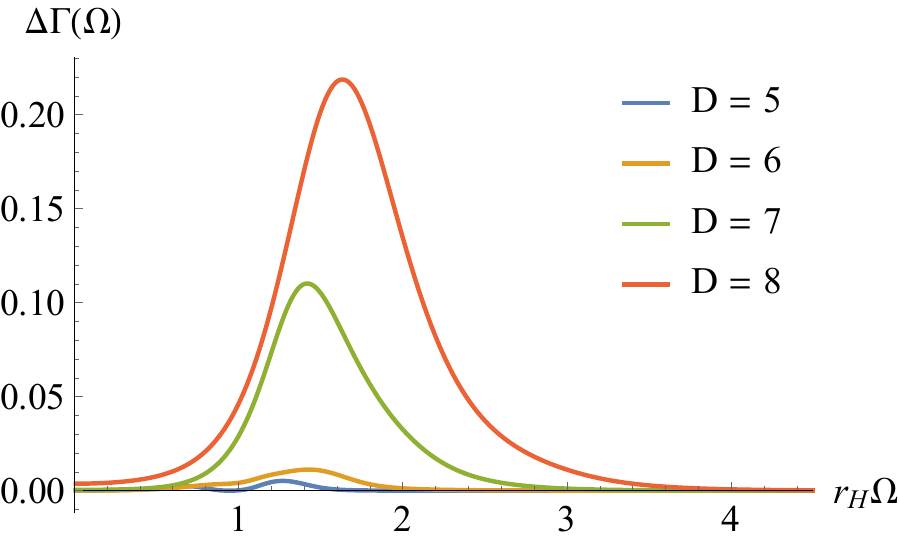}
\end{subfigure}
\caption{Left: Grey-body factors calculated by using the correspondence with the quasinormal modes for $l=2$ scalar gravitational perturbations in $D=5$ (blue), $D=6$ (yellow), $D=7$ (green), and $D=8$ (red). Right: Differences between the values obtained using the correspondence and the accurate values.} \label{sl2}
\end{figure}

\begin{figure}[h!]
\noindent\begin{subfigure}[b]{0.5\textwidth}
    \centering
    \includegraphics[scale=0.47]{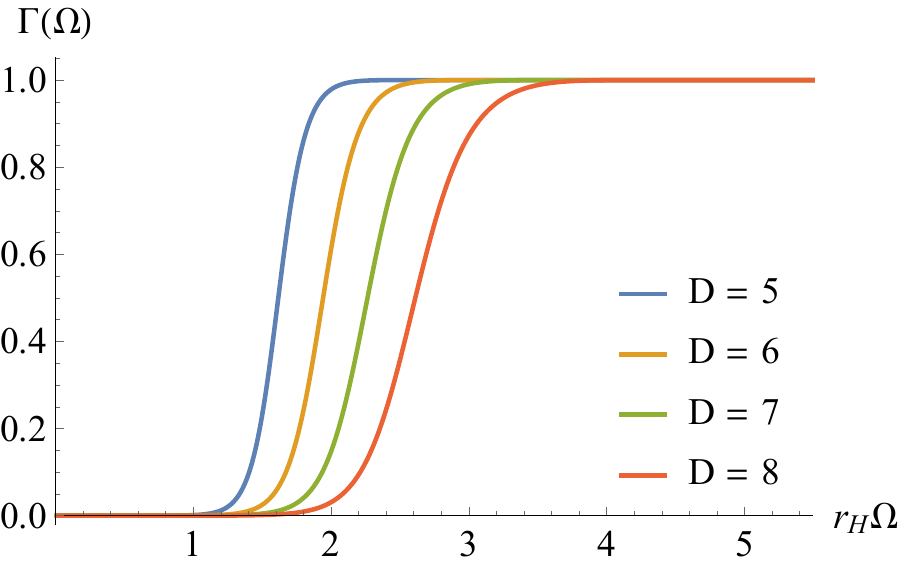}
\end{subfigure}%
\noindent\begin{subfigure}[b]{0.5\textwidth}
    \centering
    \includegraphics[scale=0.47]{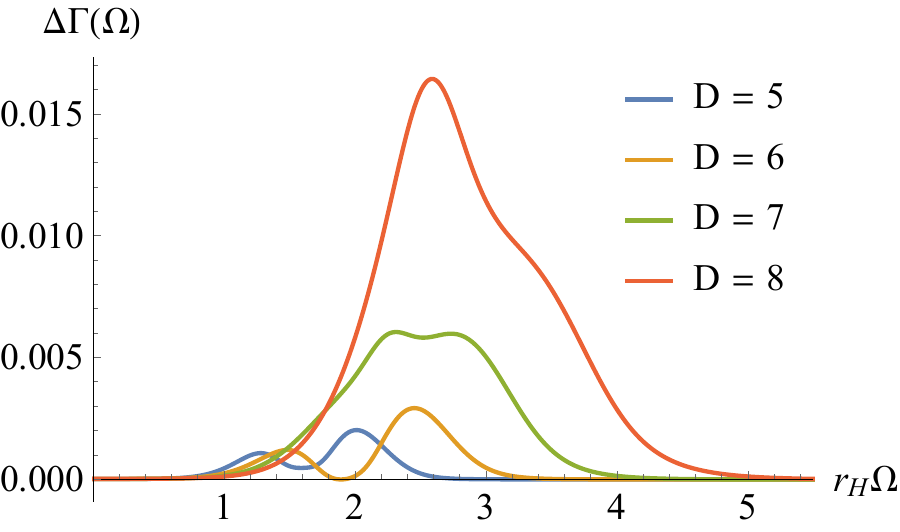}
\end{subfigure}
\caption{Left: Grey-body factors calculated by using the correspondence with the quasinormal modes for $l=3$ scalar gravitational perturbations in $D=5$ (blue), $D=6$ (yellow), $D=7$ (green), and $D=8$ (red). Right: Differences between the values obtained using the correspondence and the accurate values.} \label{sl3}
\end{figure}

\begin{figure}[h!] 
\noindent\begin{subfigure}[b]{0.5\textwidth}
    \centering
    \includegraphics[scale=0.47]{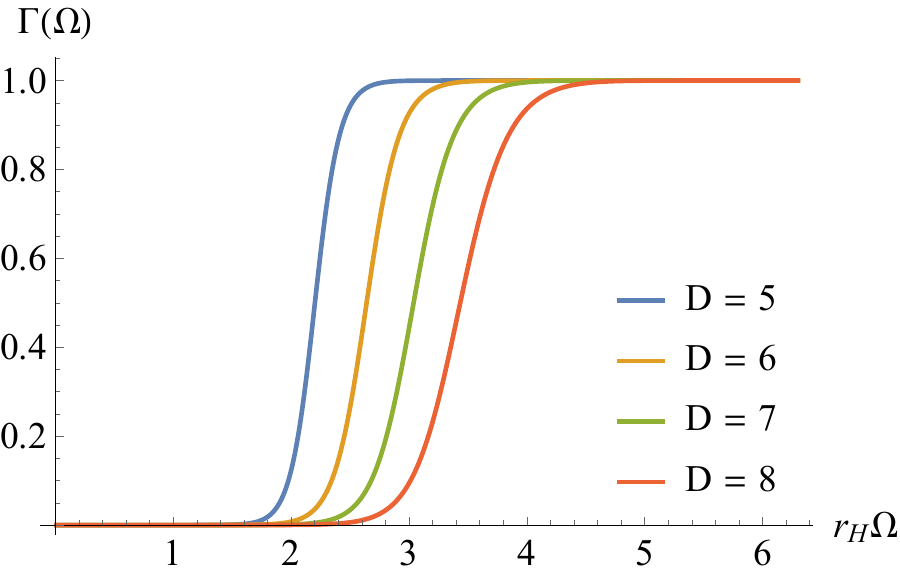}
\end{subfigure}%
\noindent\begin{subfigure}[b]{0.5\textwidth}
    \centering
    \includegraphics[scale=0.47]{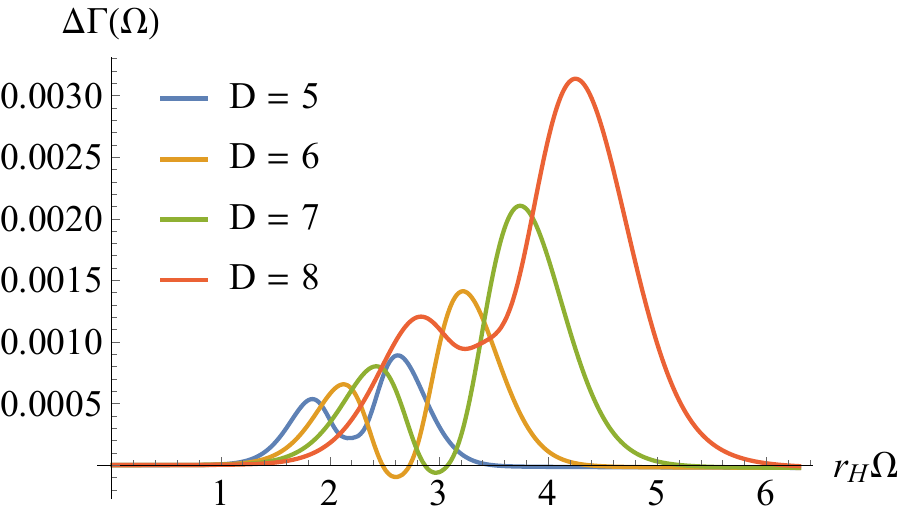}
\end{subfigure}
\caption{Left: Grey-body factors calculated by the correspondence with the quasinormal modes for $l=4$ scalar gravitational perturbations in $D=5$ (blue), $D=6$ (yellow), $D=7$ (green), and $D=8$ (red). Right: Differences between the values obtained using the correspondence and the accurate values.} \label{sl4}
\end{figure}

Furthermore, we plotted the curves of $\Gamma(\Omega)$ and $\Delta \Gamma(\Omega) $ for different $D$ values while $l$ was fixed. \autoref{sl2} shows that the correspondence for the $l=2$ mode of scalar-type perturbations starts to fail from $D=7$. As observed in \autoref{sl3} and \autoref{sl4}, the grey-body factors obtained from the correspondence achieved good accuracy for $l=3,4$ in $4<D \le8$. Further, $\Delta \Gamma(\Omega)$ increases as the number of dimensions increases, indicating a decrease in the precision of the correspondence. This is because the higher $D$, the lower the accuracy of the WKB formula used to construct the correspondence \cite{Konoplya:2003ii,Konoplya:2019hlu}.

\section{Correspondence for vector gravitational perturbations}
We now discuss the results for vector-type gravitational perturbations. We tested the correspondence at low multipole numbers for $4<D \le8$. The wave equation for vector-type perturbations contains additional singular points within the unit circle centered at $\rho=0$ in dimensions $D > 9$, corresponding to the complex roots of the metric function $f(r) $. As we present the results up to $D=8$ here, such singular points do not appear. Therefore, we applied the original continued fraction method to compute the quasinormal modes. The results of the fundamental mode and the first overtone in $D=6,7,8$ are shown in \autoref{qnmV}.

\begin{table}[h!]
\centering
\begin{tabular}{
|>{\centering\arraybackslash}p{0.5cm}|
 >{\centering\arraybackslash}p{3.5cm}|
 >{\centering\arraybackslash}p{3.5cm}|
 >{\centering\arraybackslash}p{3.5cm}| }
 
\hline\hline
\multicolumn{4}{|c|}{\textbf{$D=6$}} \\
\hline\hline

$n$ & $l=2$ & $l=3$ & $l=4$ \\
\hline
$0$   & $1.524660 - 0.474124 i$ & $2.188071 - 0.467877 i$ & $2.823705 - 0.472522 i
$\\
 $1$   & $1.155552 - 1.515257 i$ & $1.969576 - 1.431429 i$ & $2.662975 - 1.435973 i$\\
\hline\hline

\multicolumn{4}{|c|}{\textbf{$D=7$}} \\
\hline\hline

$n$ & $l=2$ & $l=3$ & $l=4$ \\
\hline
$0$   & $1.934455-0.612313i$ & $2.637397-0.594633i$ & $3.324882 -0.594512i
$\\
 $1$   & $1.382718-1.970301i$ & $2.276699 - 1.815880i$ & $3.059624-1.803542i$\\
\hline\hline

\multicolumn{4}{|c|}{\textbf{$D=8$}} \\
\hline\hline

$n$ & $l=2$ & $l=3$ & $l=4$ \\
\hline
$0$   & $2.358830-0.738736 i$ & $3.086560-0.714730 i$ & $3.809069-0.709316 i
$\\
 $1$   & $1.592418-2.338689 i$ & $2.564206-2.169676 i$ & $3.419343-2.141444i$\\
\hline\hline

\end{tabular}
     \caption{Quasinormal frequency $r_H \omega_n$ for vector gravitational perturbations in $D=6,7,8$} 
\label{qnmV}
\end{table}

\begin{figure}[h!] 
\noindent\begin{subfigure}[b]{0.5\textwidth}
    \centering
    \includegraphics[scale=0.38]{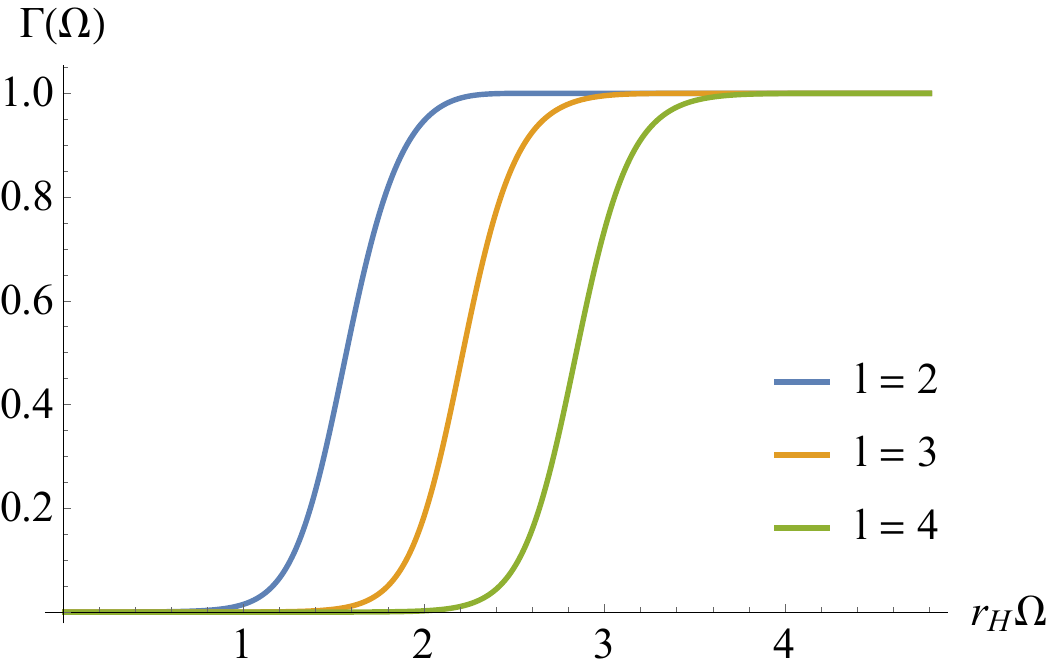}
\end{subfigure}%
\noindent\begin{subfigure}[b]{0.5\textwidth}
    \centering
    \includegraphics[scale=0.4]{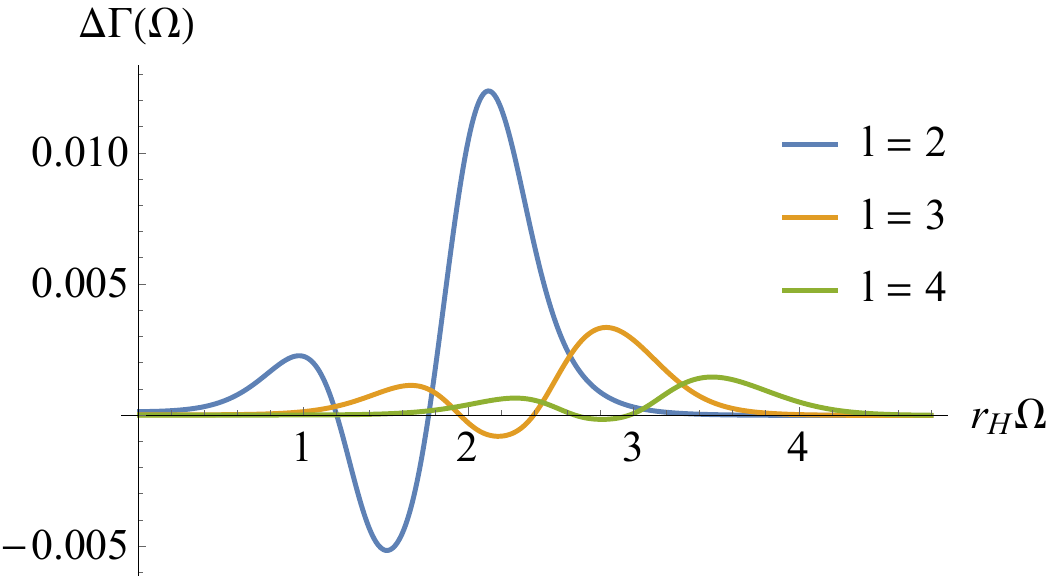}
\end{subfigure}
\caption{Left: Grey-body factors obtained by the correspondence with the quasinormal modes for vector gravitational perturbations for $l=2$ (blue), $l=3$ (yellow), and $l=4$ (green) in $D=6$. Right: Differences between the grey-body factors calculated by using the correspondence and the numerical method.} \label{fig6v}
\end{figure}

\begin{figure}[h!] 
\noindent\begin{subfigure}[b]{0.5\textwidth}
    \centering
    \includegraphics[scale=0.36]{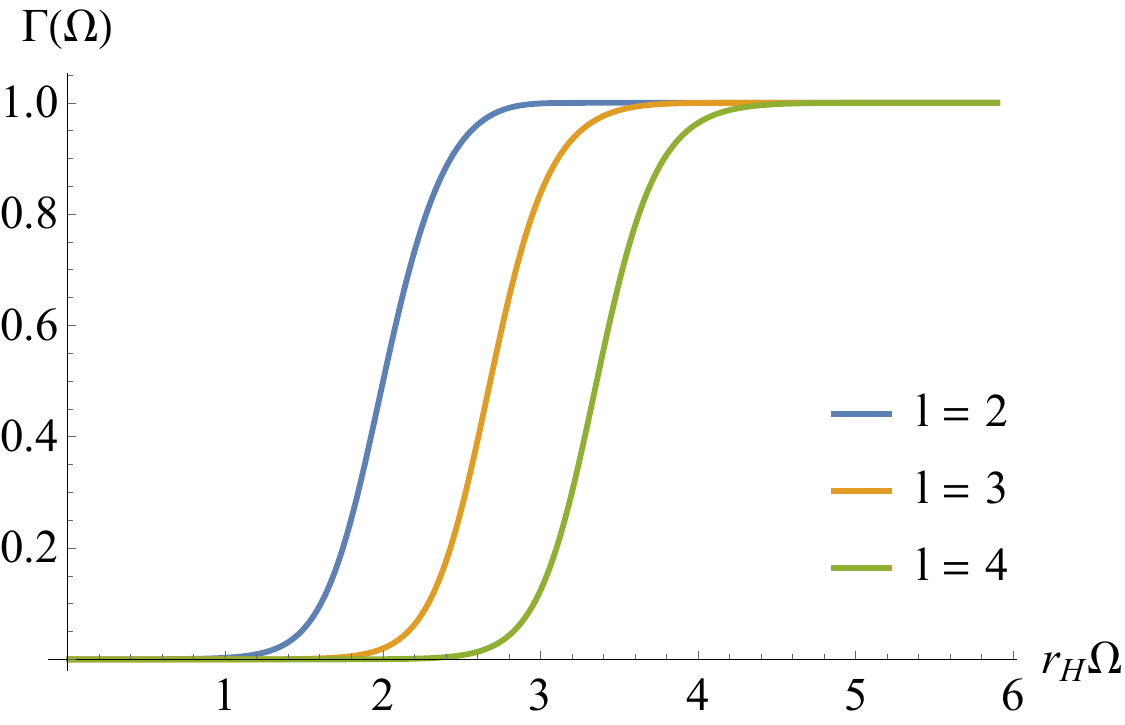}
\end{subfigure}%
\noindent\begin{subfigure}[b]{0.5\textwidth}
    \centering
    \includegraphics[scale=0.37]{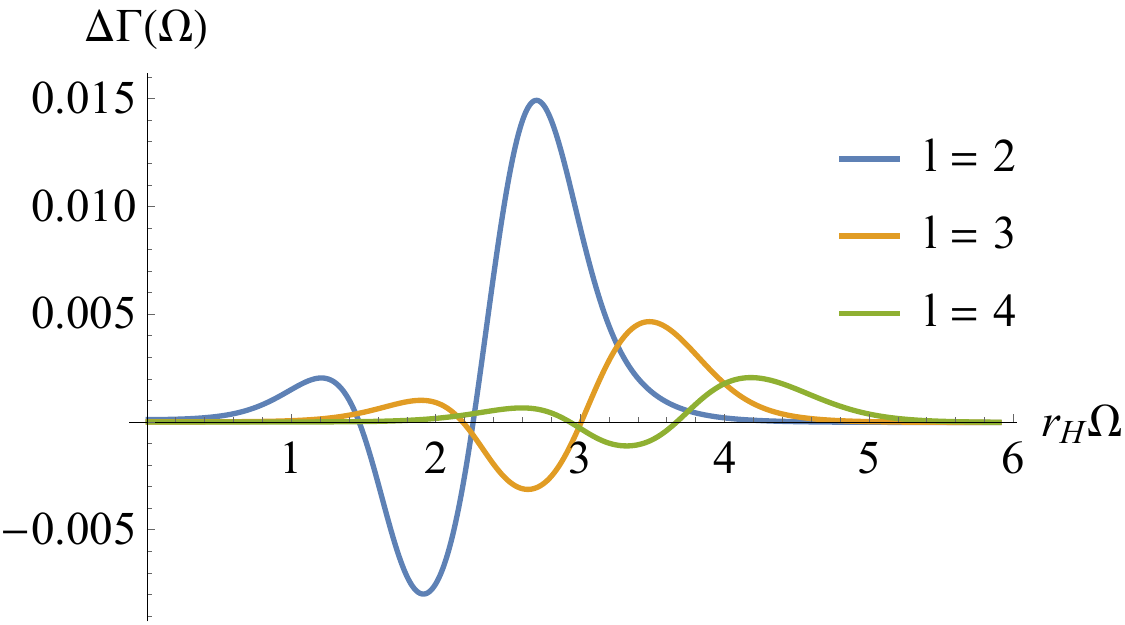}
\end{subfigure}
\caption{Left: Grey-body factors obtained by the correspondence with the quasinormal modes for vector gravitational perturbations for $l=2$ (blue), $l=3$ (yellow), and $l=4$ (green) in $D=7$. Right: Differences between the grey-body factors calculated by using the correspondence and the numerical method.} \label{fig7v}
\end{figure}

\begin{figure}[h!] 
\noindent\begin{subfigure}[b]{0.5\textwidth}
    \centering
    \includegraphics[scale=0.36]{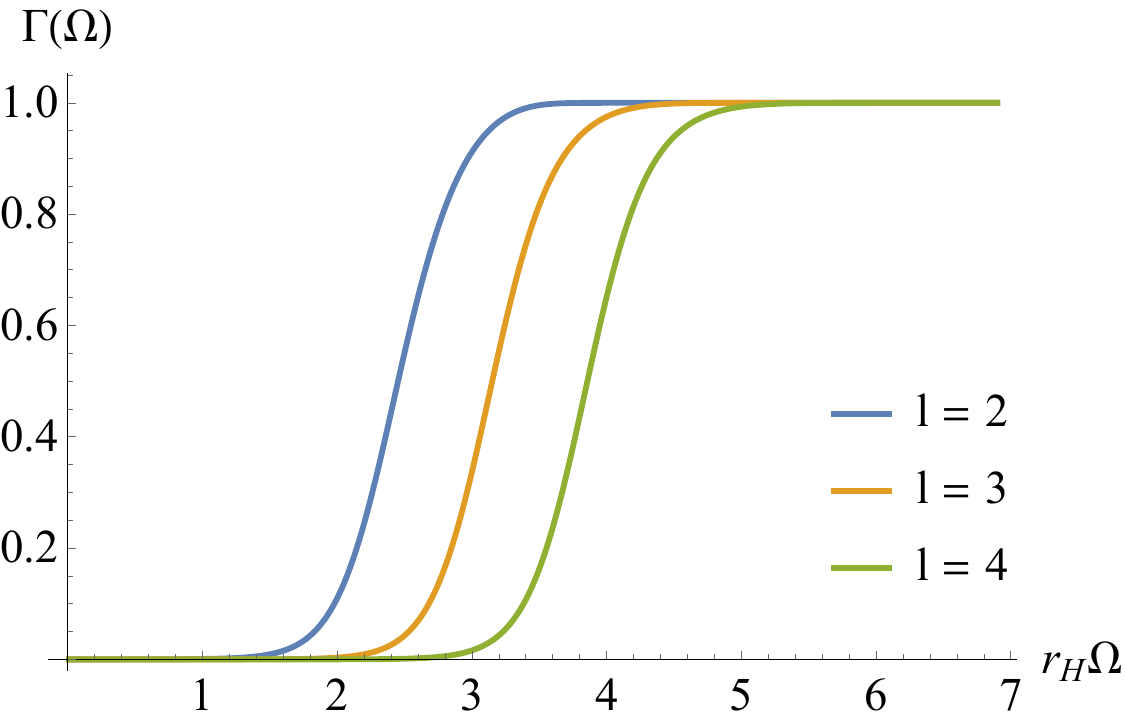}
\end{subfigure}%
\noindent\begin{subfigure}[b]{0.5\textwidth}
    \centering
    \includegraphics[scale=0.37]{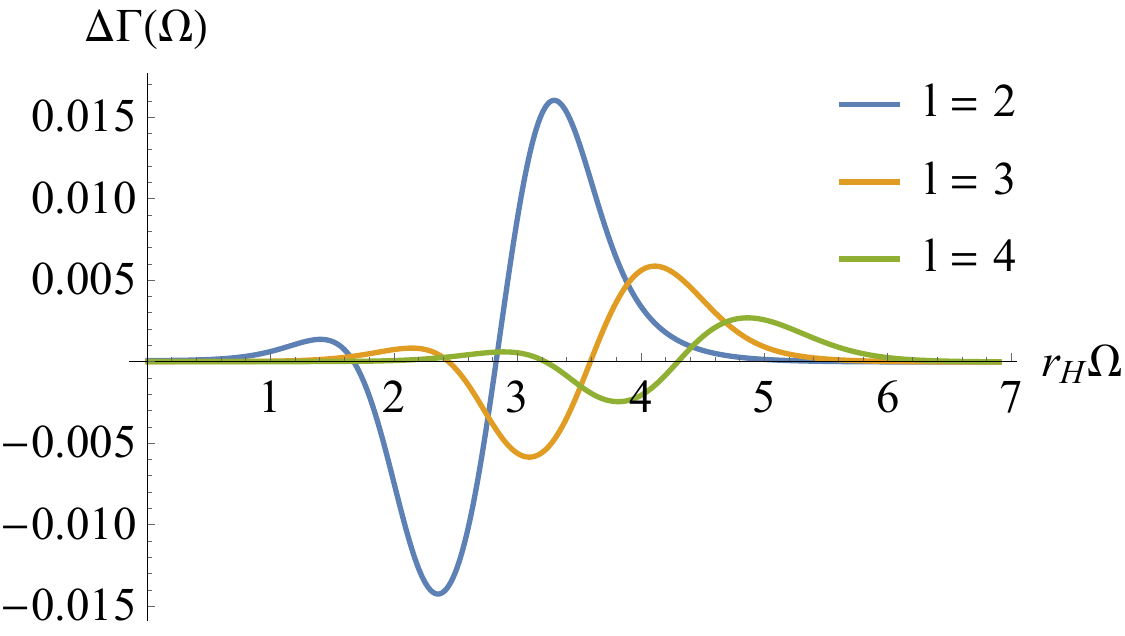}
\end{subfigure}
\caption{Left: Grey-body factors obtained by the correspondence with the quasinormal modes for vector gravitational perturbations for $l=2$ (blue), $l=3$ (yellow), and $l=4$ (green) in $D=8$. Right: Differences between the grey-body factors calculated by using the correspondence and the numerical method.} \label{fig8v}
\end{figure}

We substituted the quasinormal frequencies $\omega_0$ and $\omega_1$ into the formula \eqref{corr} and analytically obtained the grey-body factor for each case. Furthermore, we verified the accuracy of the results by calculating the difference between the values of the grey-body factor found from the correspondence and those obtained by numerical computation. For vector-type perturbations in five dimensions, the approximate values of the grey-body factors obtained by correspondence with the quasinormal modes exhibited high precision for $l=2,3,4$, as shown in Figure 3 of \cite{Han:2025cal}. Here, we show the results for six, seven, and eight dimensions in \autoref{fig6v}--\autoref{fig8v}.

The graphs reveal small differences $\Delta \Gamma(\Omega)$ between the grey-body factors obtained from the analytical formula and the numerical method at low multipole numbers. This suggests that the correspondence between the quasinormal modes and grey-body factors is valid for vector gravitational perturbations in higher dimensions. As $l$ increases, $\Delta\Gamma(\Omega)$ decreases; thus, the accuracy of the correspondence tends to increase.

\begin{figure}[h!] 
\noindent\begin{subfigure}[b]{0.33\textwidth}
    \centering
    \includegraphics[scale=0.55]{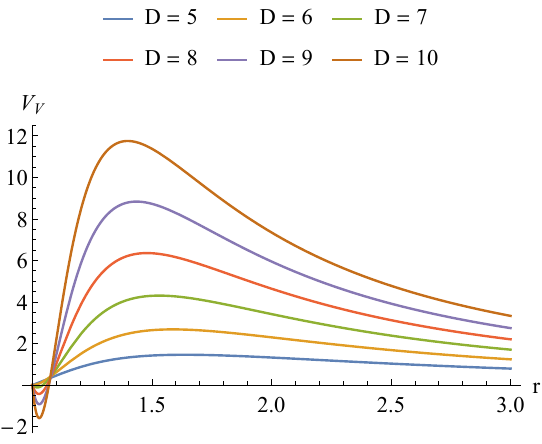}
\end{subfigure}%
\noindent\begin{subfigure}[b]{0.33\textwidth}
    \centering
    \includegraphics[scale=0.55]{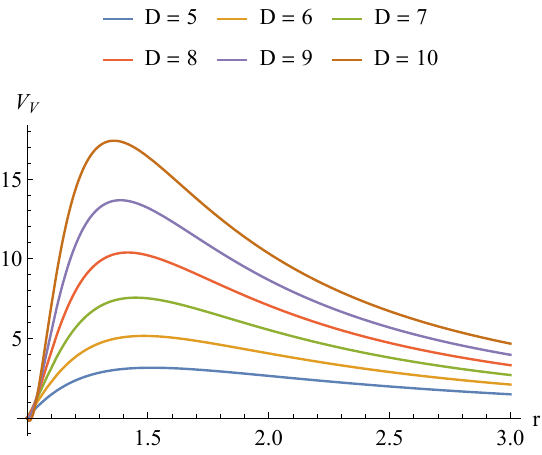}
\end{subfigure}%
\noindent\begin{subfigure}[b]{0.33\textwidth}
    \centering
    \includegraphics[scale=0.55]{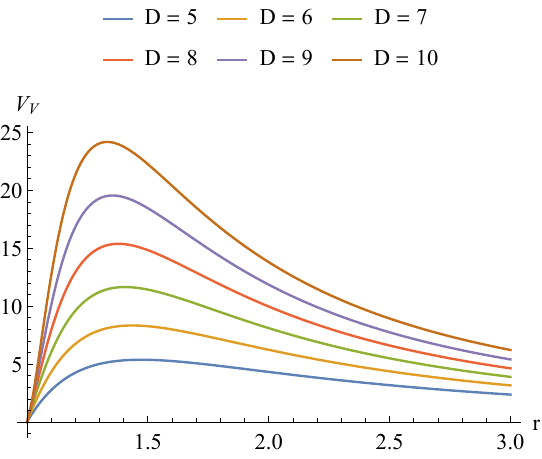}
\end{subfigure}%
\caption{Effective potentials for vector-type gravitational perturbations of Schwarzschild--Tangherlini black holes ($r_H=1$) for $l=2$ (left), $l=3$ (center), and $l=4$ (right).} \label{VVp}
\end{figure}

To analyze the origin of the validity of the correspondence and its behavior for higher $D$, we plotted the effective potential $V_V$ \eqref{VV} for the vector-type perturbations of the Schwarzschild--Tangherlini background in \autoref{VVp}. In contrast to scalar-type perturbations, vector-type perturbations do not admit multiple barriers for all multipole numbers $l \ge 2$ in higher dimensions. These forms of potential underlie the validity of the connection between the quasinormal modes and grey-body factors. Hence, this correspondence is expected to be applicable even for higher $D$ than those considered in this work.

\begin{figure}[h!] 
\noindent\begin{subfigure}[b]{0.5\textwidth}
    \centering
    \includegraphics[scale=0.43]{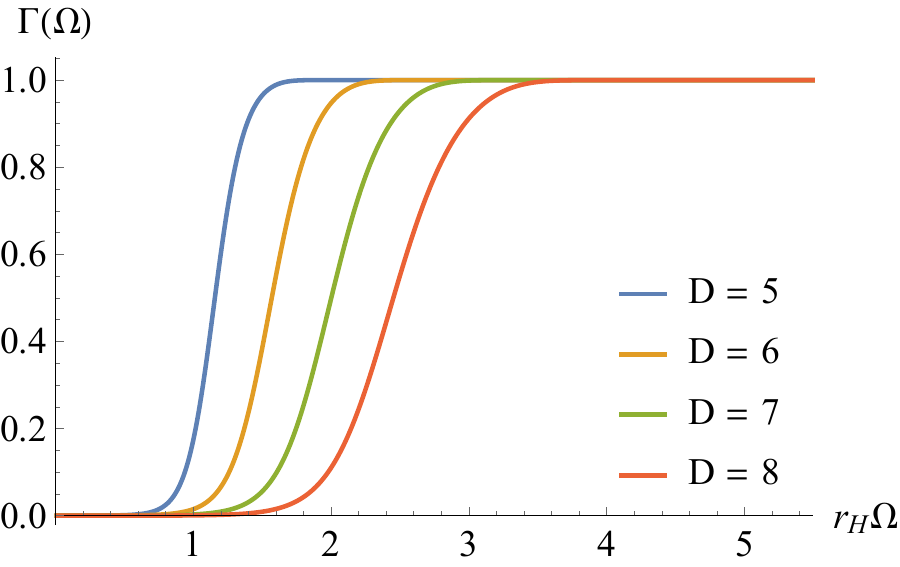}
\end{subfigure}%
\noindent\begin{subfigure}[b]{0.5\textwidth}
    \centering
    \includegraphics[scale=0.42]{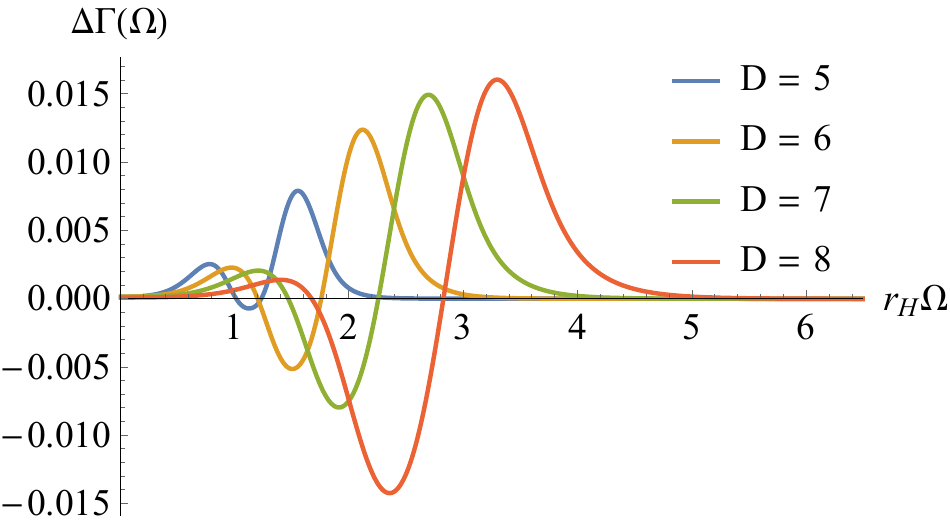}
\end{subfigure}
\caption{Left: Grey-body factors calculated by using the correspondence with the quasinormal modes for $l=2$ vector gravitational perturbations in $D=5$ (blue), $D=6$ (yellow), $D=7$ (green), and $D=8$ (red). Right: Differences between the values obtained using the correspondence and the accurate values.} \label{vl2}
\end{figure}

\begin{figure}[H] 
\noindent\begin{subfigure}[b]{0.5\textwidth}
    \centering
    \includegraphics[scale=0.43]{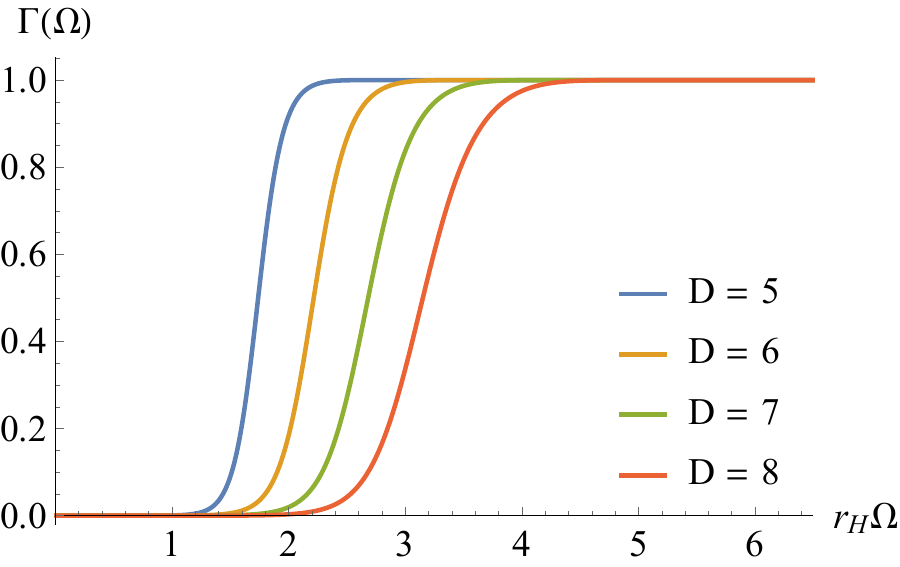}
\end{subfigure}%
\noindent\begin{subfigure}[b]{0.5\textwidth}
    \centering
    \includegraphics[scale=0.42]{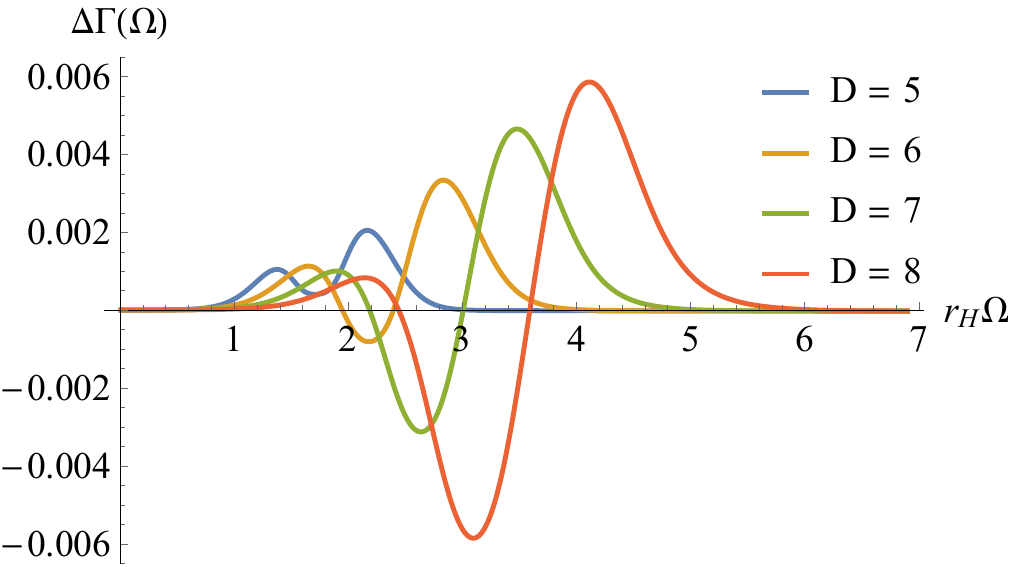}
\end{subfigure}
\caption{Left: Grey-body factors calculated by using the correspondence with the quasinormal modes for $l=3$ vector gravitational perturbations in $D=5$ (blue), $D=6$ (yellow), $D=7$ (green), and $D=8$ (red). Right: Differences between the values obtained using the correspondence and the accurate values.} \label{vl3}
\end{figure}

\begin{figure}[h!] 
\noindent\begin{subfigure}[b]{0.5\textwidth}
    \centering
    \includegraphics[scale=0.43]{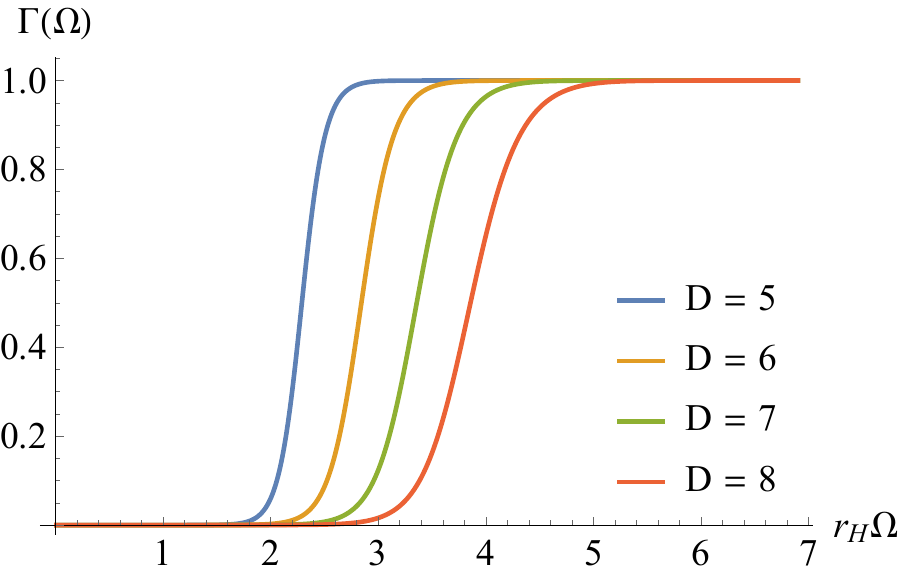}
\end{subfigure}%
\noindent\begin{subfigure}[b]{0.5\textwidth}
    \centering
    \includegraphics[scale=0.42]{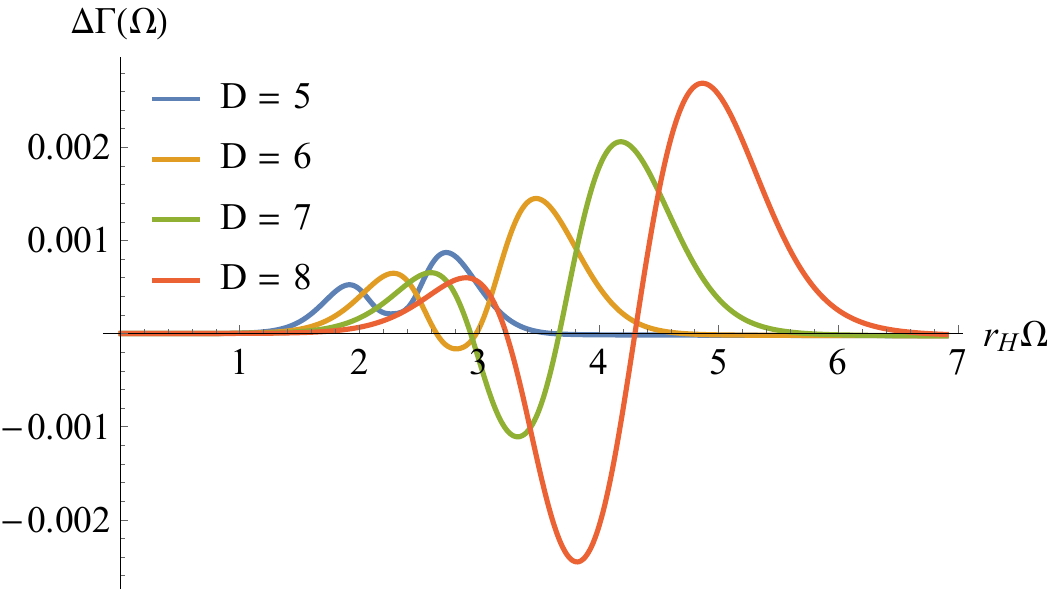}
\end{subfigure}
\caption{Left: Grey-body factors calculated by using the correspondence with the quasinormal modes for $l=4$ vector gravitational perturbations in $D=5$ (blue), $D=6$ (yellow), $D=7$ (green), and $D=8$ (red). Right: Differences between the values obtained using the correspondence and the accurate values.} \label{vl4}
\end{figure}

We also present graphs for each value of $l$ in \autoref{vl2}--\autoref{vl4}. For vector gravitational perturbations, the precision of the correspondence in all cases tends to decrease with an increasing number of dimensions.

\section{Correspondence for tensor gravitational perturbations}

Finally, we examined the correspondence for tensor-type gravitational perturbations. Similar to the case of vector-type perturbations, the wave equation for tensor-type perturbations contains no additional singularities within the convergence radius of the Frobenius series \eqref{frob} in $D\le9$. Thus, in this investigation, for $4 < D \le 8$, the quasinormal modes can be extracted from Eq. \eqref{n0} with an infinite continued fraction without the need to introduce midpoints. \autoref{qnmT} provides the frequencies of the fundamental mode and the first overtone for tensor-type perturbations in $D=6,7,8$.

\begin{table}[h!]
\centering
\begin{tabular}{
|>{\centering\arraybackslash}p{0.5cm}|
 >{\centering\arraybackslash}p{3.5cm}|
 >{\centering\arraybackslash}p{3.5cm}|
 >{\centering\arraybackslash}p{3.5cm}| }
 
\hline\hline
\multicolumn{4}{|c|}{\textbf{$D=6$}} \\
\hline\hline

$n$ & $l=2$ & $l=3$ & $l=4$ \\
\hline
$0$   & $2.011534 - 0.501938 i$ & $2.579088 - 0.498872 i$ & $3.147805 - 0.497326 i
$\\
 $1$   & $1.786041 - 1.558792 i$ & $2.402351 - 1.527257 i$ & $3.002869 - 1.512010 i$\\
\hline\hline

\multicolumn{4}{|c|}{\textbf{$D=7$}} \\
\hline\hline

$n$ & $l=2$ & $l=3$ & $l=4$ \\
\hline
$0$   & $2.496780 - 0.631882 i$ & $3.114073 - 0.627615 i$ & $3.732388 - 0.625330 i
$\\
 $1$   & $2.137269-1.961169 i$ & $2.830631 - 1.921398 i$ & $3.498188-1.901644i$\\
\hline\hline

\multicolumn{4}{|c|}{\textbf{$D=8$}} \\
\hline\hline

$n$ & $l=2$ & $l=3$ & $l=4$ \\
\hline
$0$   & $2.974689-0.750560 i$ & $3.629848-0.745301 i$ & $4.285781-0.742343 i
$\\
 $1$   & $2.457466-2.313075 i$ & $3.221393-2.272610 i$ & $3.946624-2.251894i$\\
\hline\hline

\end{tabular}
    \caption{Quasinormal frequency $r_H \omega_n$ for tensor gravitational perturbations in $D=6,7,8$} 
\label{qnmT}
\end{table}

\begin{figure}[h!] 
\noindent\begin{subfigure}[b]{0.5\textwidth}
    \centering
    \includegraphics[scale=0.45]{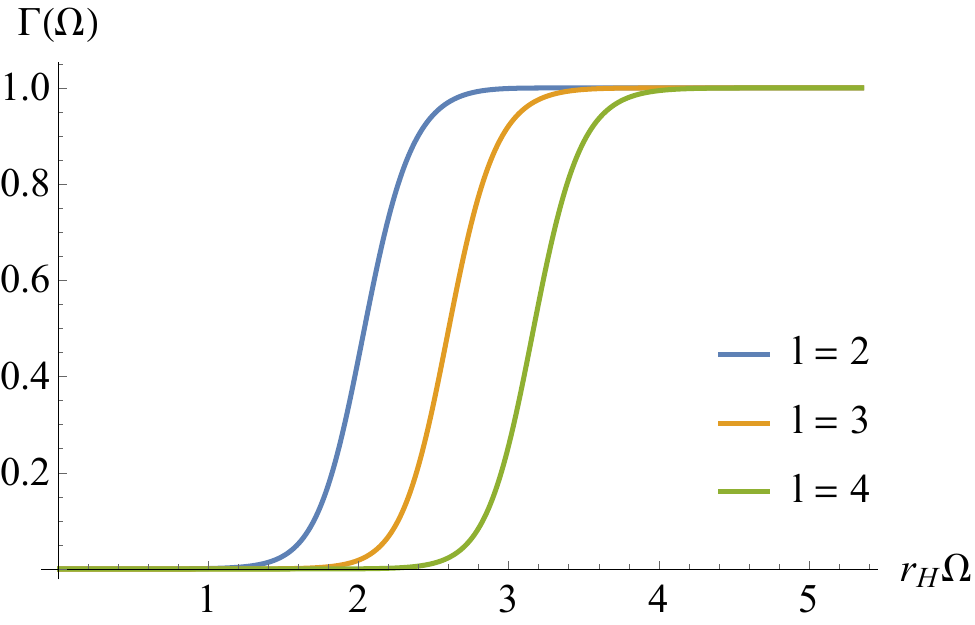}
\end{subfigure}%
\noindent\begin{subfigure}[b]{0.5\textwidth}
    \centering
    \includegraphics[scale=0.45]{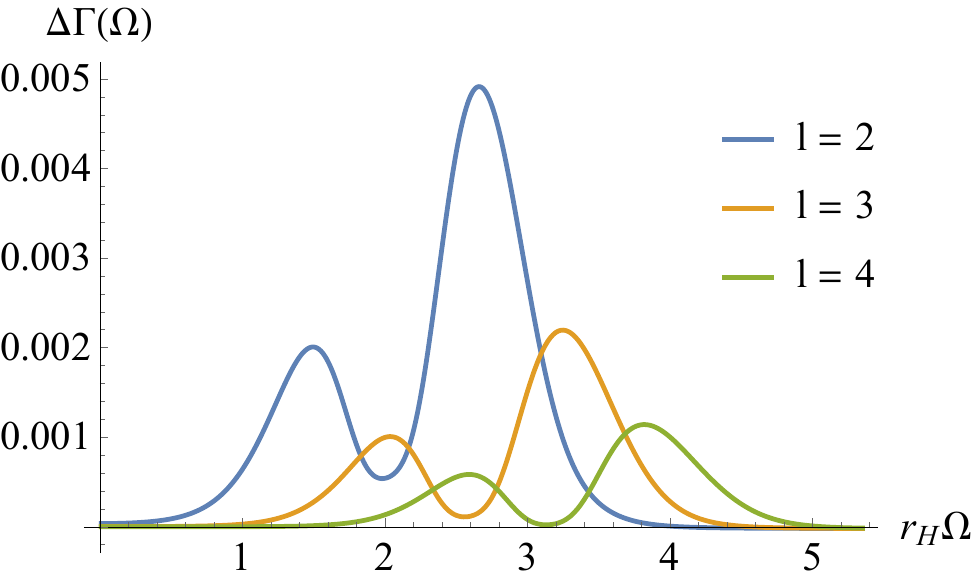}
\end{subfigure}
\caption{Left: Grey-body factors obtained by using the correspondence with the quasinormal modes for tensor gravitational perturbations for $l=2$ (blue), $l=3$ (yellow), and $l=4$ (green) in $D=6$. Right: Differences between the grey-body factors calculated by using the correspondence and the numerical method.} \label{fig6t}
\end{figure}

\begin{figure}[h!] 
\noindent\begin{subfigure}[b]{0.5\textwidth}
    \centering
    \includegraphics[scale=0.37]{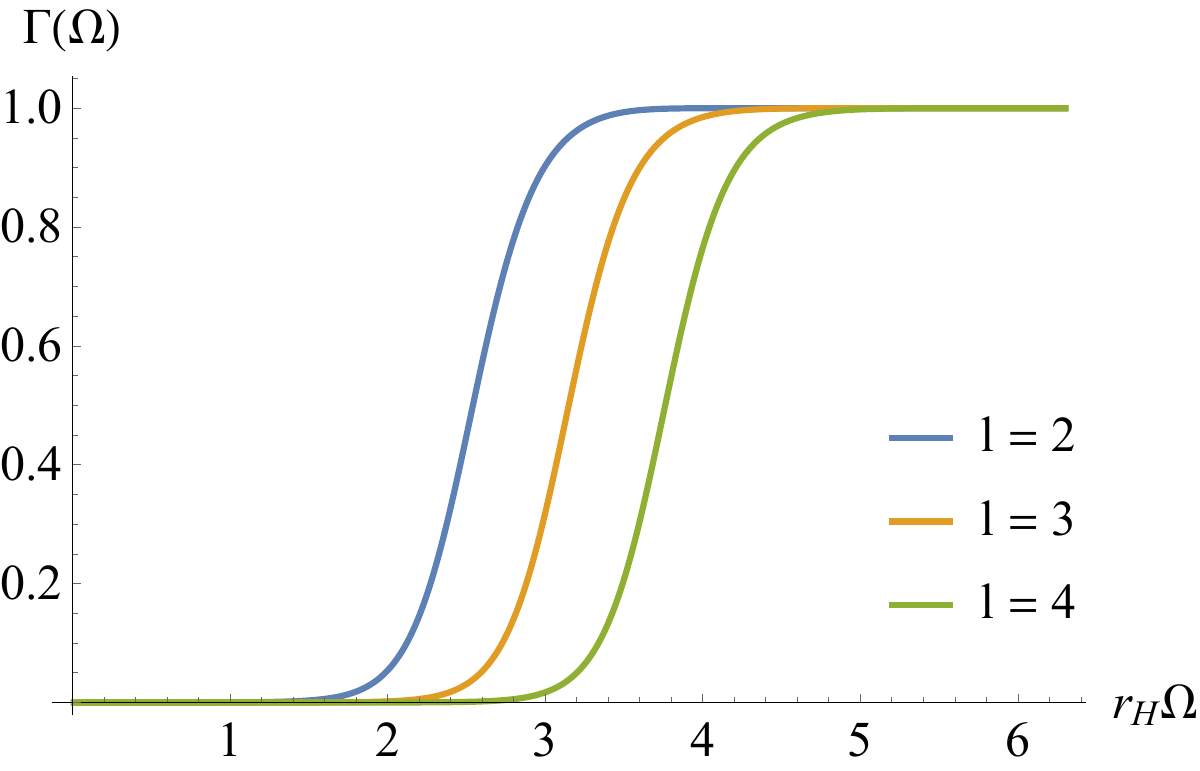}
\end{subfigure}%
\noindent\begin{subfigure}[b]{0.5\textwidth}
    \centering
    \includegraphics[scale=0.37]{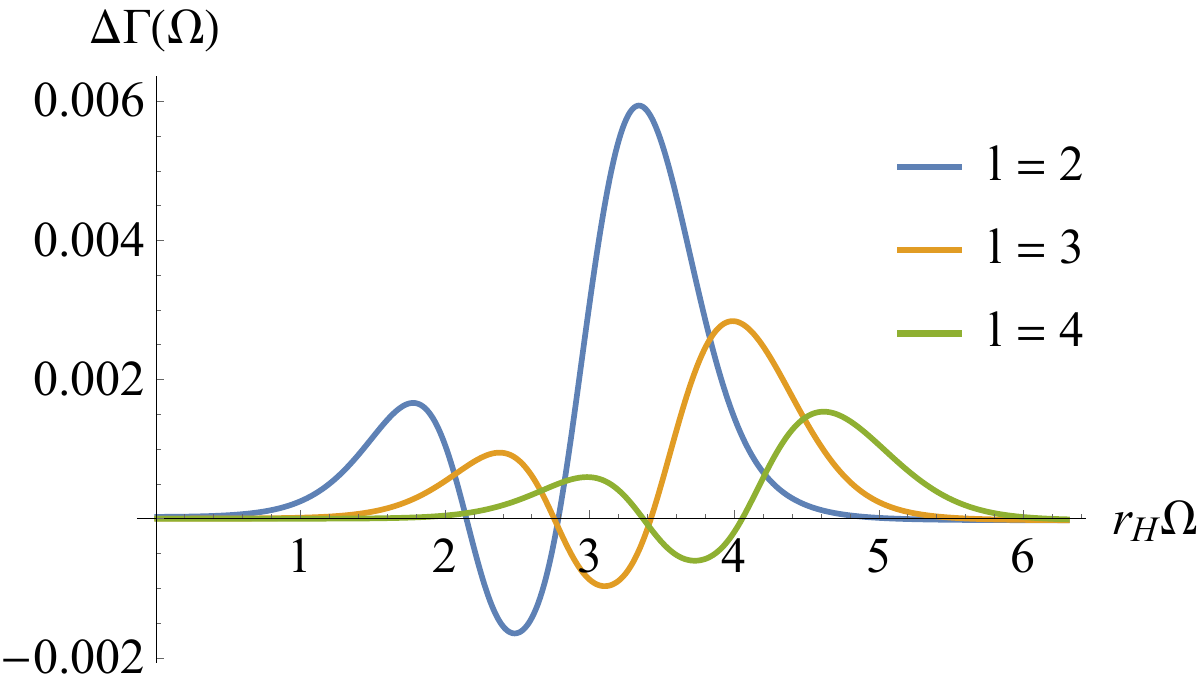}
\end{subfigure}
\caption{Left: Grey-body factors obtained by using the correspondence with the quasinormal modes for tensor gravitational perturbations for $l=2$ (blue), $l=3$ (yellow), and $l=4$ (green) in $D=7$. Right: Differences between the grey-body factors calculated by using the correspondence and the numerical method.} \label{fig7t}
\end{figure}

\begin{figure}[h!] 
\noindent\begin{subfigure}[b]{0.5\textwidth}
    \centering
    \includegraphics[scale=0.38]{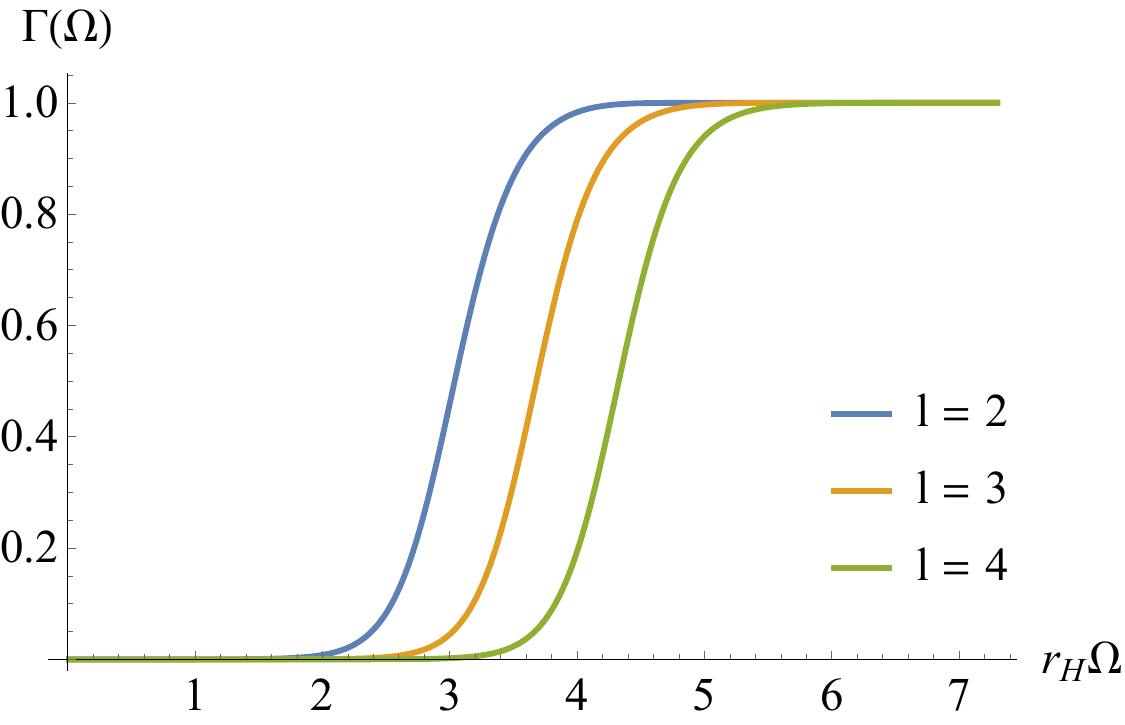}
\end{subfigure}%
\noindent\begin{subfigure}[b]{0.5\textwidth}
    \centering
    \includegraphics[scale=0.38]{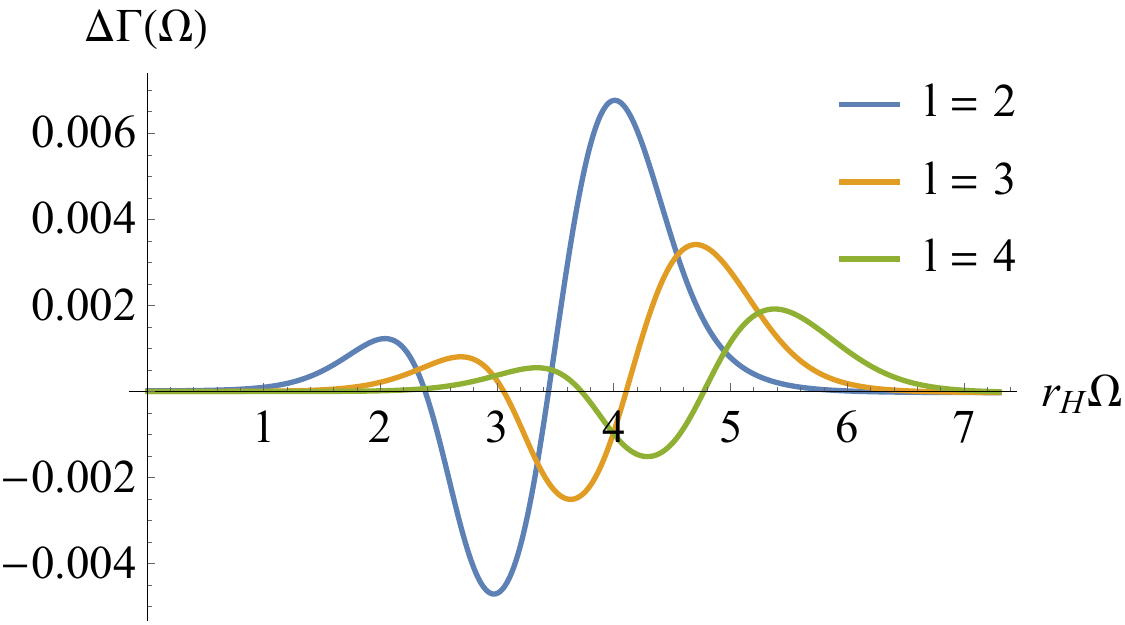}
\end{subfigure}
\caption{Left: Grey-body factors obtained by using the correspondence with the quasinormal modes for tensor gravitational perturbations for $l=2$ (blue), $l=3$ (yellow), and $l=4$ (green) in $D=8$. Right: Differences between the grey-body factors calculated by using the correspondence and the numerical method.} \label{fig8t}
\end{figure}

Following the methods described in the previous sections, we substituted the numerical results for the quasinormal modes into Eqs. \eqref{corr} to determine the grey-body factors. Figure 4 in \cite{Han:2025cal} shows that for tensor gravitational perturbations of a five-dimensional black hole, the grey-body factors can be calculated with high accuracy through their correspondence with the quasinormal modes beyond the eikonal limit. For a larger number of dimensions, we plotted the grey-body factors $\Gamma(\Omega)$ obtained through the correspondence and show the difference $\Delta \Gamma(\Omega)$ between them and the numerically computed values in \autoref{fig6t}--\autoref{fig8t}.

\begin{figure}[h!] 
\noindent\begin{subfigure}[b]{0.33\textwidth}
    \centering
    \includegraphics[scale=0.55]{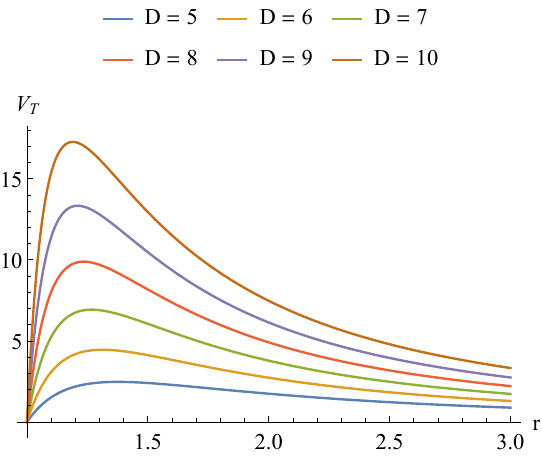}
\end{subfigure}%
\noindent\begin{subfigure}[b]{0.33\textwidth}
    \centering
    \includegraphics[scale=0.55]{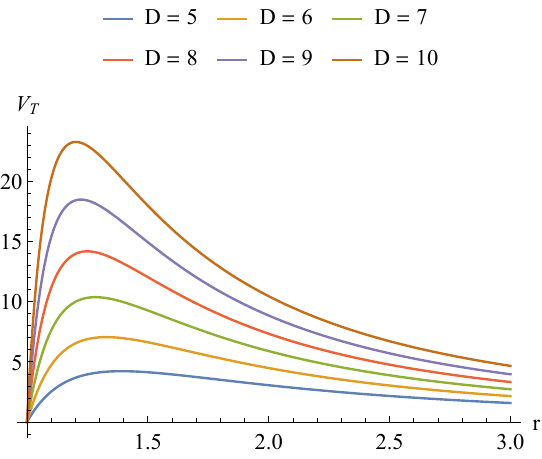}
\end{subfigure}%
\noindent\begin{subfigure}[b]{0.33\textwidth}
    \centering
    \includegraphics[scale=0.55]{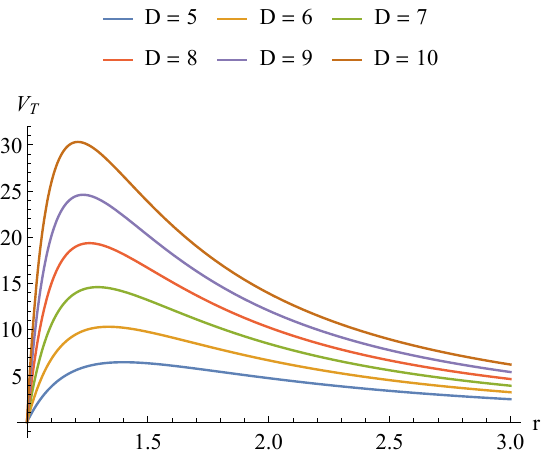}
\end{subfigure}%
\caption{Effective potentials for tensor-type gravitational perturbations of Schwarzschild--Tangherlini black holes ($r_H=1$) for $l=2$ (left), $l=3$ (center), and $l=4$ (right).} \label{VTp} 
\end{figure}

These results demonstrate that the correspondence between the quasinormal modes and grey-body factors for tensor-type perturbations yields high accuracy in all cases. As shown in \autoref{VTp}, the effective potential $V_T$ \eqref{VT} for the tensor type is positive definite with a single barrier for all higher dimensions and multipole numbers anywhere outside the black hole. The remarkable validity of this correspondence is attributed to these potentials belonging to the class for which the WKB method performs well.

\begin{figure}[h!] 
\noindent\begin{subfigure}[b]{0.5\textwidth}
    \centering
    \includegraphics[scale=0.29]{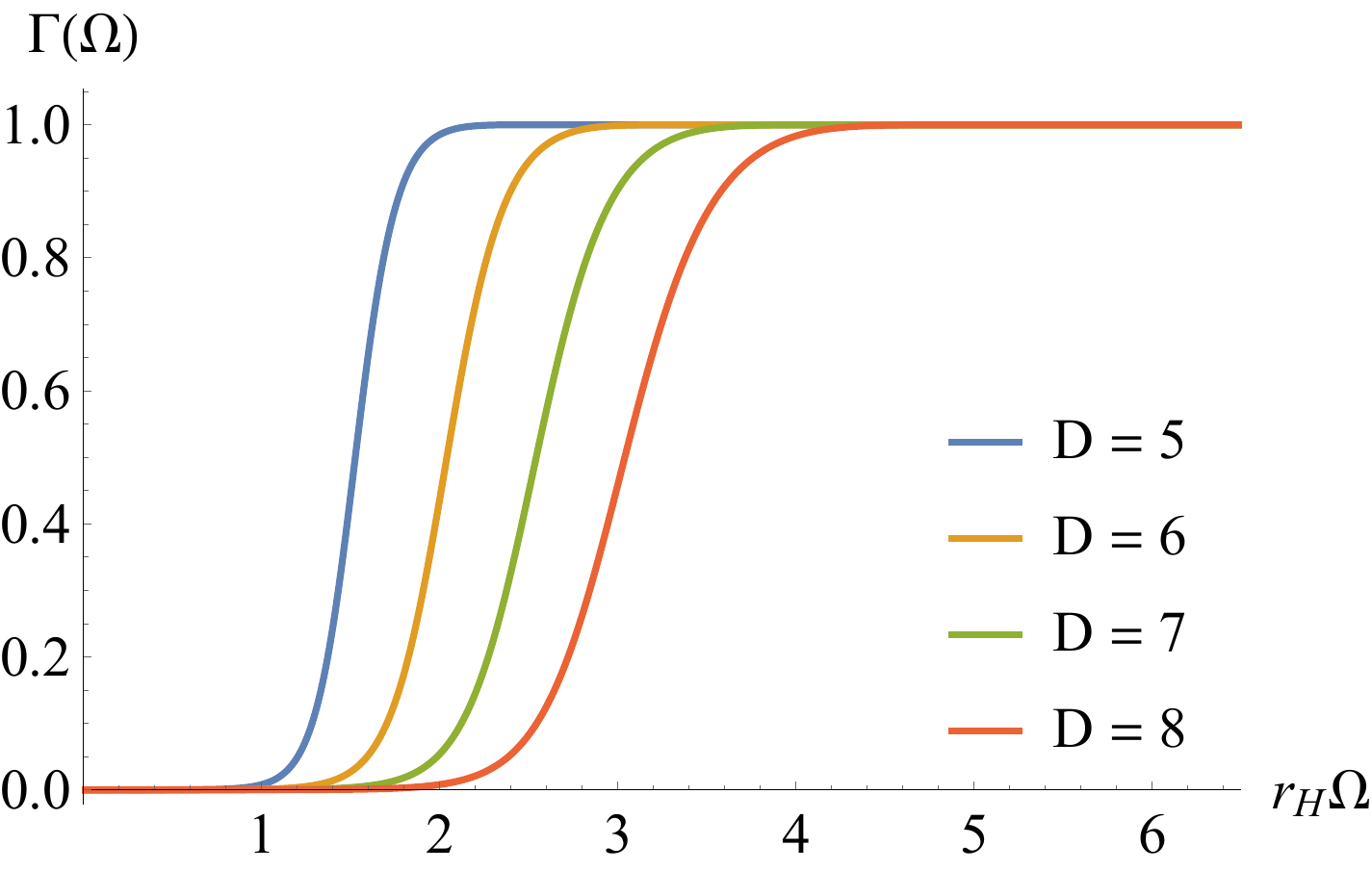}
\end{subfigure}%
\noindent\begin{subfigure}[b]{0.5\textwidth}
    \centering
    \includegraphics[scale=0.28]{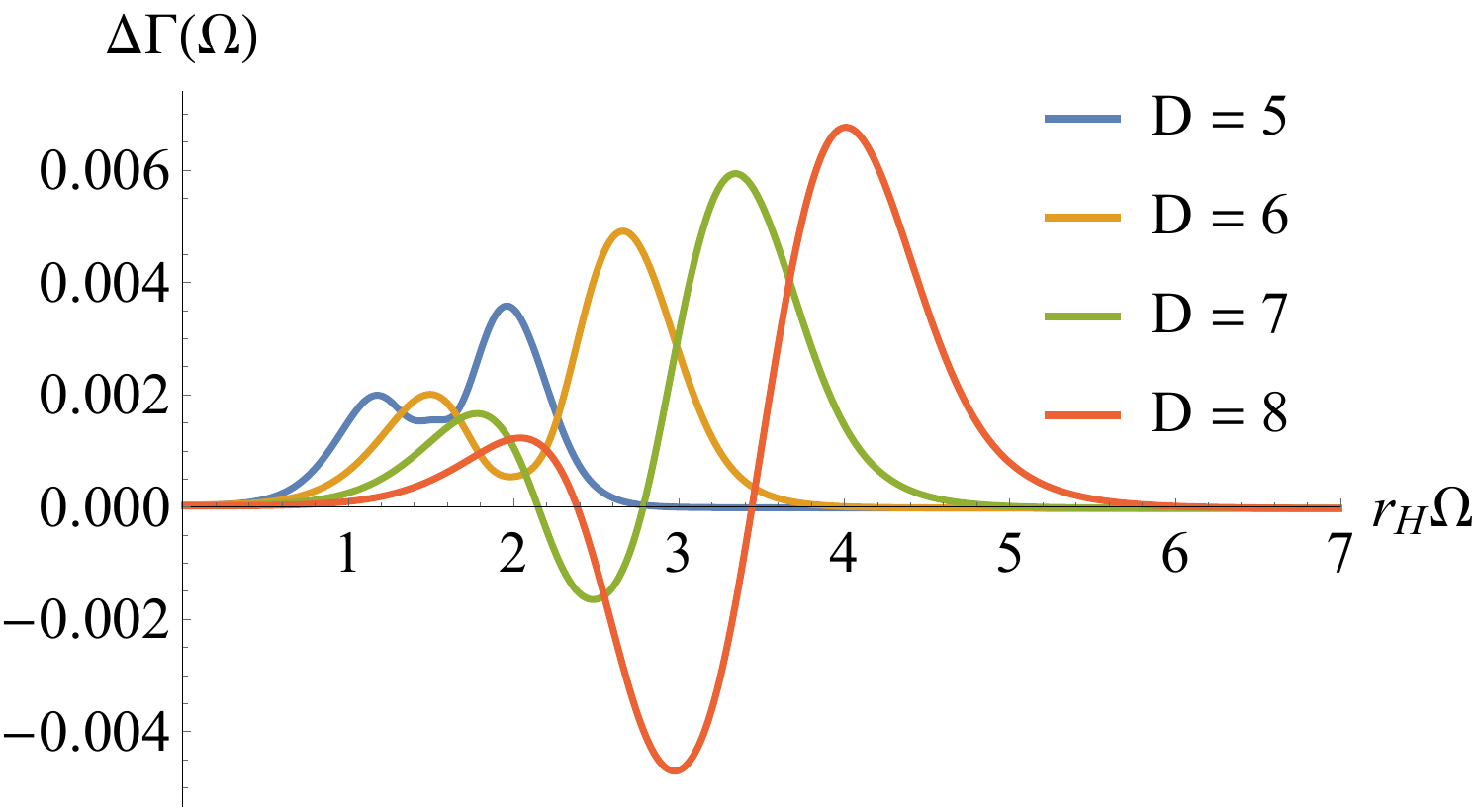}
\end{subfigure}
\caption{Left: Grey-body factors calculated by using the correspondence with the quasinormal modes for $l=2$ tensor gravitational perturbations in $D=5$ (blue), $D=6$ (yellow), $D=7$ (green), and $D=8$ (red). Right: Differences between the values obtained using the correspondence and the accurate values.} \label{tl2}
\end{figure}

\begin{figure}[h!] 
\noindent\begin{subfigure}[b]{0.5\textwidth}
    \centering
    \includegraphics[scale=0.29]{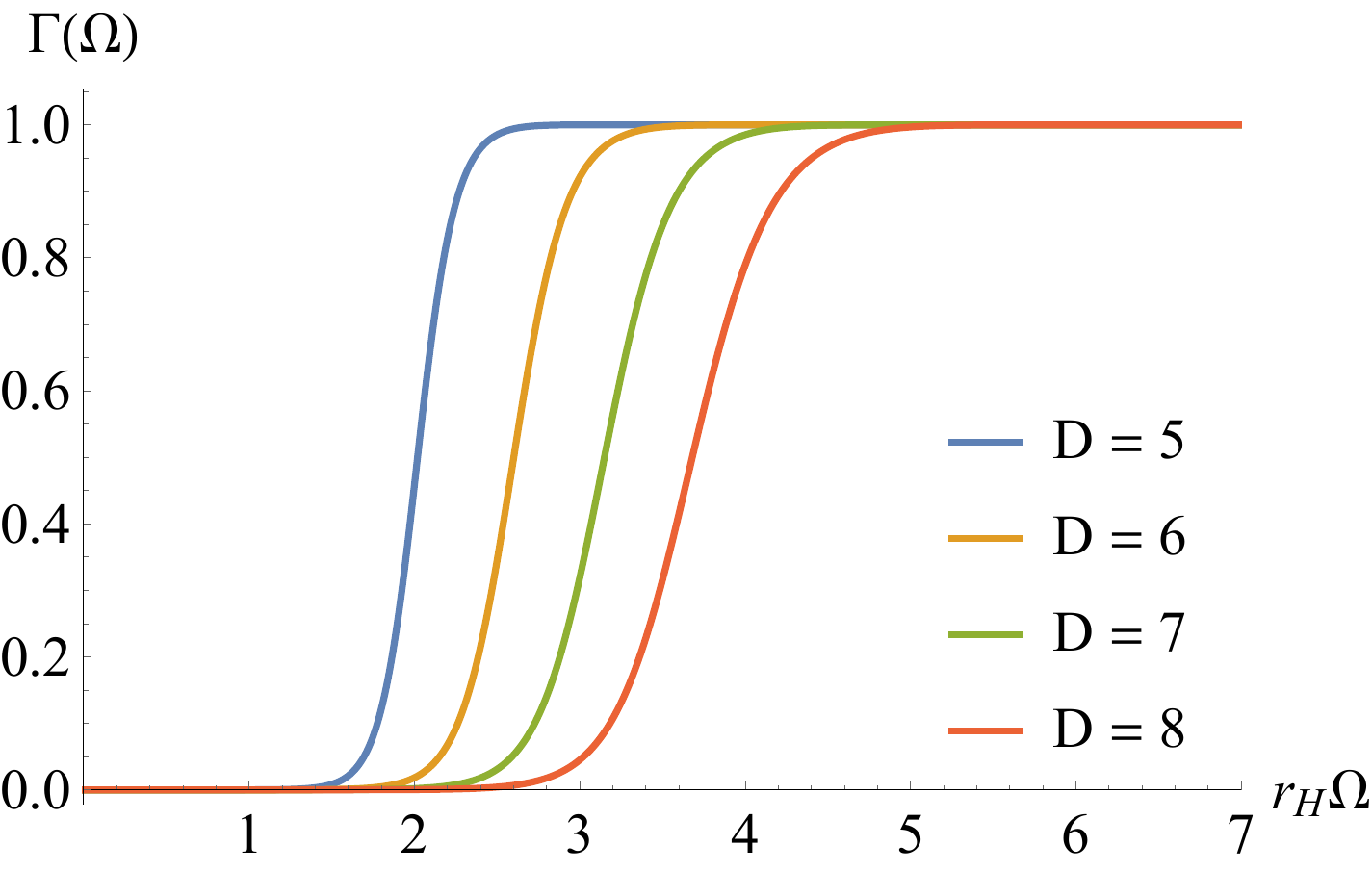}
\end{subfigure}%
\noindent\begin{subfigure}[b]{0.5\textwidth}
    \centering
    \includegraphics[scale=0.28]{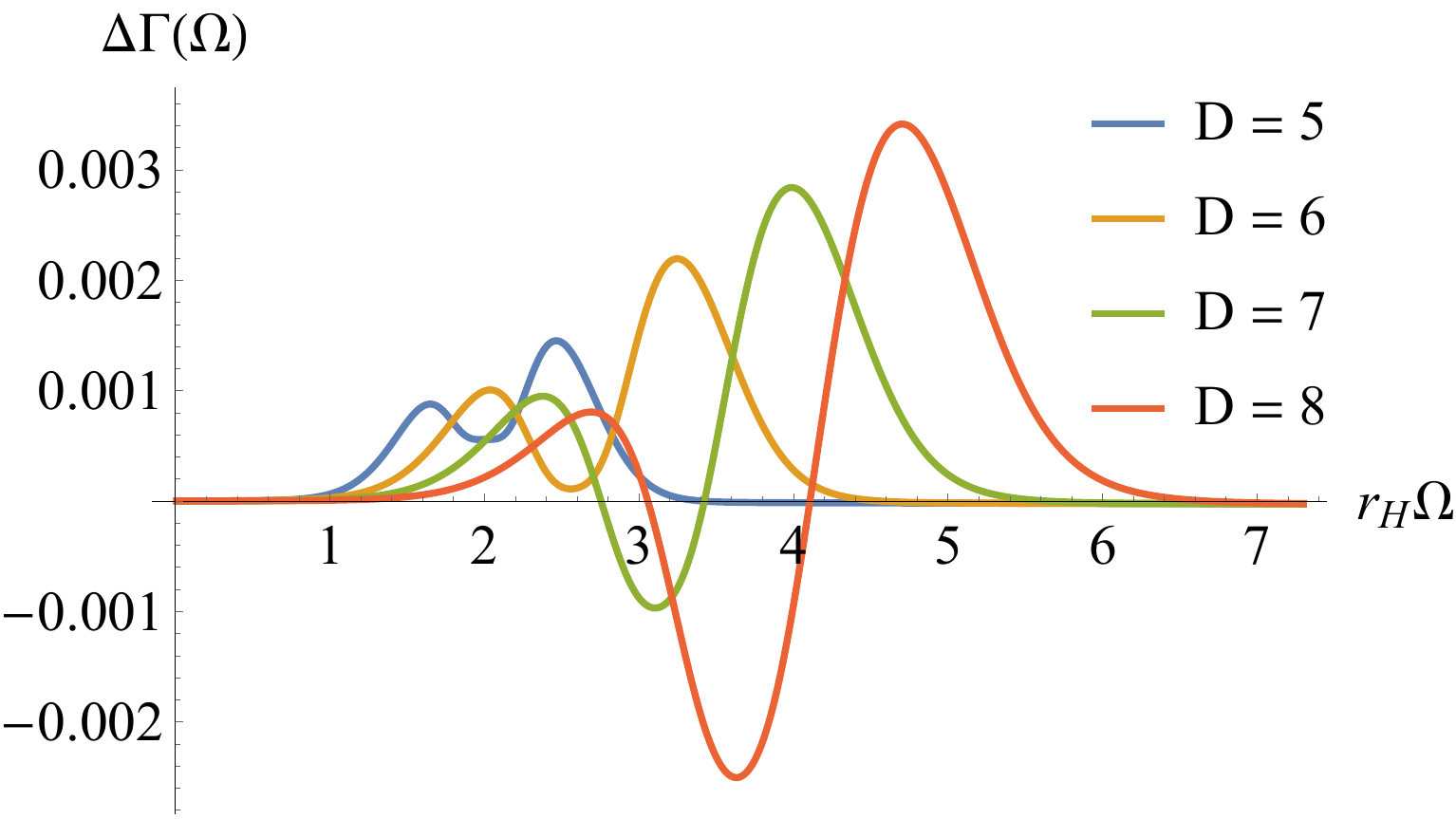}
\end{subfigure}
\caption{Left: Grey-body factors calculated by using the correspondence with the quasinormal modes for $l=3$ tensor gravitational perturbations in $D=5$ (blue), $D=6$ (yellow), $D=7$ (green), and $D=8$ (red). Right: Differences between the values obtained using the correspondence and the accurate values.} \label{tl3}
\end{figure}

\begin{figure}[h!] 
\noindent\begin{subfigure}[b]{0.5\textwidth}
    \centering
    \includegraphics[scale=0.29]{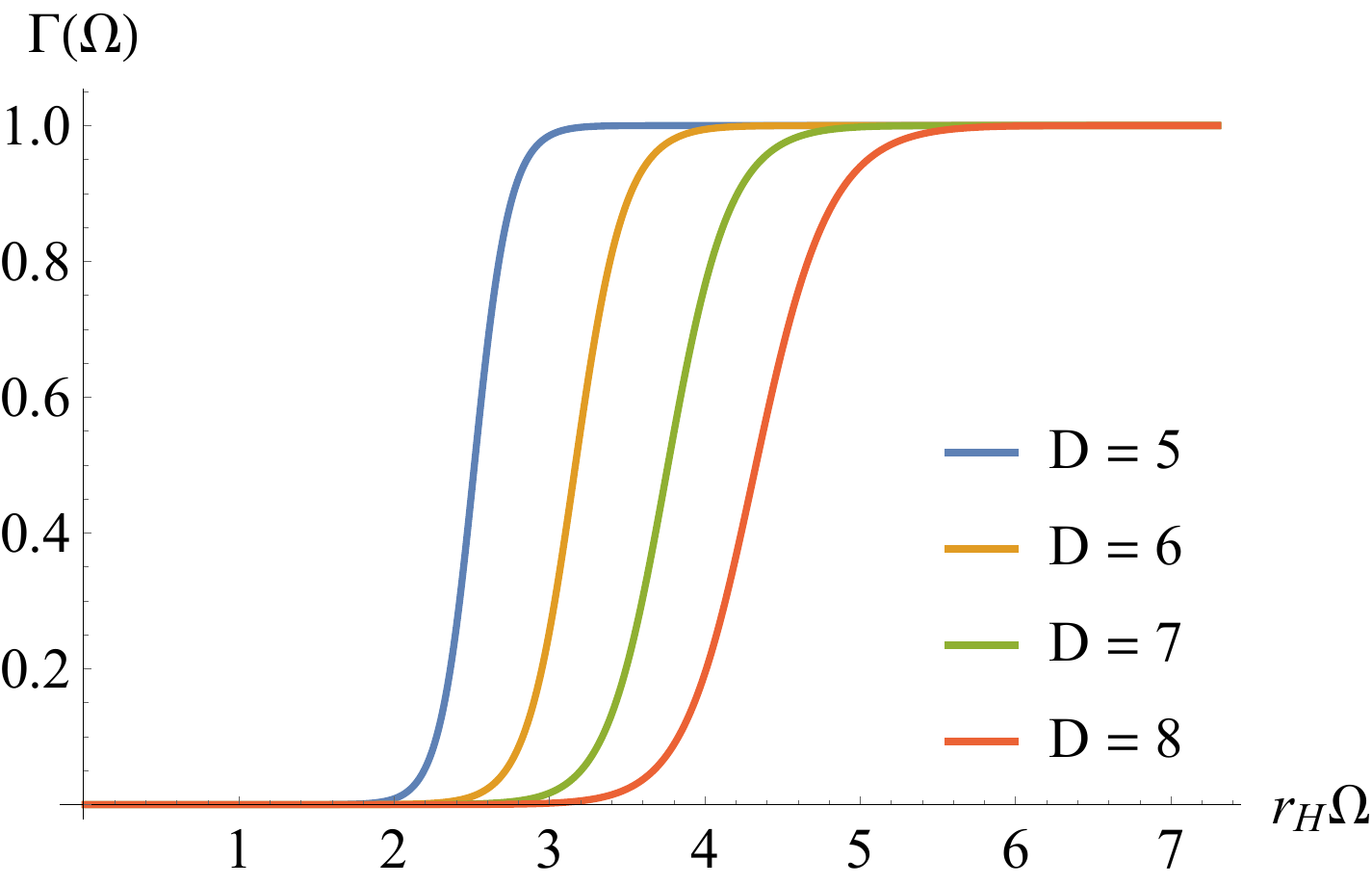}
\end{subfigure}%
\noindent\begin{subfigure}[b]{0.5\textwidth}
    \centering
    \includegraphics[scale=0.27]{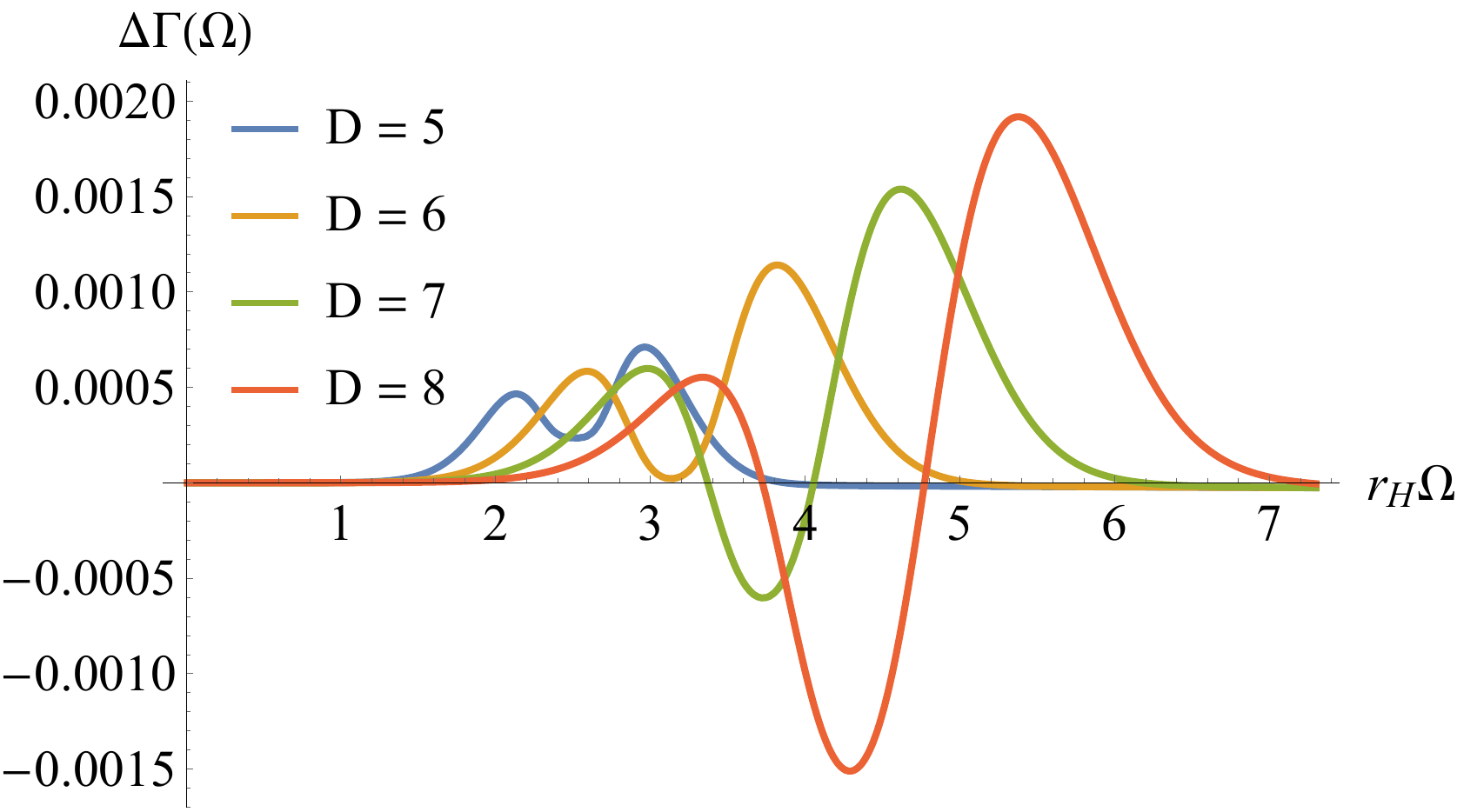}
\end{subfigure}
\caption{Left: Grey-body factors calculated by using the correspondence with the quasinormal modes for $l=4$ tensor gravitational perturbations in $D=5$ (blue), $D=6$ (yellow), $D=7$ (green), and $D=8$ (red). Right: Differences between the values obtained using the correspondence and the accurate values.} \label{tl4}
\end{figure}

The graphs in \autoref{fig6t}--\autoref{fig8t} clearly show decreases in $\Delta \Gamma(\Omega)$ with increases in $l$, and thus improvements in the accuracy of the correspondence in each dimension. Moreover, we present the plots of $\Gamma(\Omega)$ and $\Delta \Gamma(\Omega)$ for each multipole number $l=2,3,4$ in \autoref{tl2}--\autoref{tl4}. As in the cases of the other types of perturbations, the higher the dimension, the greater the difference $\Delta \Gamma(\Omega)$ because of the nature of the higher-order WKB formula. Among the three types, the tensor-type perturbation achieves the highest accuracy for the correspondence. The magnitude $|\Delta \Gamma(\Omega)|$ even for the lowest multipole number $l=2$ in $D=8$ remains below $0.007$, as shown in \autoref{tl2}, compared with the case of vector-type perturbations in \autoref{vl2}, for which the maximum value of $|\Delta \Gamma(\Omega)|$ is approximately $0.016$.

Note that the wavelike equation \eqref{waveeq} with the effective potential $V_T$ \eqref{VT} for tensor gravitational perturbations is identical to that for perturbations of massless scalar fields \cite{Konoplya:2003ii,Konoplya:2003dd}. Therefore, the results in this section suggest that the correspondence between the quasinormal mode and grey-body factor is also valid, with high accuracy for scalar field perturbations for $l \ge 2$ in all higher dimensions $D \ge 5$.

\section{Conclusions}
We examined the correspondence between the quasinormal modes and grey-body factors of spherically symmetric black hole spacetime generalized to higher dimensions $D > 4$. By evaluating the accuracy of the correspondence for the scalar, vector, and tensor types of gravitational perturbations in each dimension, we found that it breaks down in certain cases of the scalar type. This is attributed to the fact that the effective potential for scalar gravitational perturbations deviates from the class in which the WKB approach is applicable depending on the dimension and multipole numbers. We demonstrated that for vector- and tensor-type perturbations, the correspondence holds in all cases.

The relationship between the quasinormal modes and grey-body factors was originally derived for four-dimensional spherically symmetric black holes via the sixth-order WKB formula. In the eikonal regime with a large multipole number $l$, the correspondence provides exact values of the grey-body factors in terms of the fundamental quasinormal mode. It is approximate at smaller $l$, and its precision is improved by adding beyond-eikonal corrections that contain overtones. We employed Schwarzschild--Tangherlini black holes to analyze the correspondence for the three types of gravitational perturbations in higher dimensions. The eikonal regime in this context can be accurately described using the WKB method \cite{Cardoso:2008bp}. We investigated the validity of the correspondence at the lowest multipole numbers by computing the accuracy of the grey-body factors obtained by the correspondence.

To obtain the grey-body factors for scalar, vector, and tensor types of gravitational perturbations, we adopted the continued fraction method to compute the accurate quasinormal modes. Because the original continued fraction method fails in some cases for higher-dimensional black holes, we applied the integration-through-midpoints technique. The numerical results for the fundamental mode $n=0$ and the first overtone $n=1$ at $D=6,7,8$ were presented explicitly. We substituted the quasinormal frequencies into the analytical formula to determine the approximate grey-body factors for low values of $l$. Furthermore, to verify their accuracy, we calculated the differences from the reference values obtained via numerical integration of the wave equation. The magnitudes of these differences clearly indicated whether the correspondence was valid for each case.

Our previous work \cite{Han:2025cal} proved that the correspondence between the quasinormal modes and grey-body factors of Schwarzschild--Tangherlini black holes holds for all three types of gravitational perturbations in $D=5$. For higher $D$, we analyzed the results for each type of perturbation. The scalar type achieved good accuracy of the correspondence at $l=2,3,4$ in $D=6$. However, in $D\ge7$, the correspondence for the $l=2$ mode broke down because the WKB approach is inapplicable to its effective potential with multiple barriers. Consequently, we expect that the correspondence for the $l=3$ and $l=4$ modes of scalar-type perturbations will fail in $D\ge10$ and in $D\ge12$, respectively. 
The vector- and tensor-type perturbations, for which the effective potential is characterized by a single positive barrier in all cases, showed the validity of the correspondence with high precision. 
We rigorously verified that the correspondence exists only when the potential has the standard form of the WKB method. For all types of gravitational perturbations, the accuracy of the correspondence improved with increasing $l$. This tendency is consistent with the fact that it is exact in the eikonal regime. Furthermore, because the precision of the WKB formula is better for smaller $D$, the grey-body factors obtained from the correspondence also exhibited higher precision for smaller $D$. Among the three types of perturbation, the tensor type generally showed the highest accuracy. Because the perturbations of massless scalar fields are governed by the same wave equation as tensor gravitational perturbations, our results suggest a connection between quasinormal modes and grey-body factors for massless scalar perturbations in higher dimensions.

\vspace{10pt} 

\noindent{\bf Acknowledgments}

\noindent This research was supported by Basic Science Research Program through the National Research Foundation of Korea (NRF) funded by the Ministry of Education (NRF-2022R1I1A2063176) and the Dongguk University Research Fund of 2026.\\

\bibliographystyle{bibstyle}
\bibliography{ref}
\end{document}